# Learning the Chronology of Materials Discovery for Synthesizability Prediction

Junkil Park[1*] and Aron Walsh[1*]

[1] Department of Materials, Imperial College London, London SW7 2AZ, U.K.

* j.park1@imperial.ac.uk and a.walsh@imperial.ac.uk

## Abstract

High-throughput computational and generative approaches are proposing candidate materials far faster than they can be synthesized, making it increasingly important to identify which hypothetical materials are likely to be experimentally realizable. Positive-unlabeled (PU) learning has emerged as a representative framework for synthesizability prediction, yet existing approaches treat experimentally reported materials as a static collection, overlooking the historical trajectory of materials discovery and the diverse practical factors that have driven its progress. Here we introduce ChronoSyn, a chronology-aware machine learning model that jointly learns synthesizability and the relative reporting order of known materials through a shared composition representation. Incorporating discovery chronology provides additional supervision beyond conventional synthesizability labels, improving prediction performance and enabling ChronoSyn to identify experimentally realizable compositions more effectively than baseline models. Applied to approximately 50 million hypothetical compositions, ChronoSyn identifies chemically coherent high-scoring families that include experimentally reported compositions unseen during training. These results highlight the chronology of materials discovery as an informative learning signal and suggest that the path along which known materials were discovered can point the way to those that will be discovered in the future.

# Introduction

The discovery of functional materials is central to the advancement of materials science yet remains a major challenge. The chemically accessible materials space is enormous, and recent advances in high-throughput computation[1-3] and generative artificial intelligence[4-8] are further accelerating the rate at which candidate materials can be proposed. Nevertheless, experimental synthesis and characterization remain comparatively slow and resource-intensive, creating a growing imbalance between the rapidly expanding number of computationally proposed candidates and the limited capacity for experimental validation. As this imbalance continues to grow, the ability to identify and prioritize regions of materials space that are more likely to be experimentally realizable becomes increasingly important in modern materials discovery.

Earlier approaches to assessing materials synthesizability have relied on heuristic chemical rules[9,10] and thermodynamic descriptors[11,12], but their applicability is limited by the complex factors governing experimental synthesis. In recent years, a range of data-driven approaches have been developed to predict the synthesizability of materials[13-18], among which positive-unlabeled (PU) learning[19,20] has emerged as a representative framework. In this approach, experimentally reported materials are treated as positive data (P), while unreported hypothetical materials are treated as unlabeled data (U). This formulation is well suited to materials synthesizability prediction given the inherently temporal nature of the task, because hypothetical candidates are not necessarily true negatives and may include materials that have not yet been synthesized but could be experimentally realized in the future. For example, Jang et al.[17] demonstrated that PU-learning methods can capture aspects of synthesizability beyond thermodynamic stability and can practically guide experimental exploration, resulting in the discovery of a new quaternary oxide phase.

Although the PU formulation is premised on the temporal evolution of synthesizability, existing study[13-17] treat the reported materials as an unordered collection, without considering how their discoveries relate to one another over time. Yet materials discovery follows an evolving historical trajectory. Materials databases are not the product of uniform or random exploration of chemical space[21], but rather the cumulative record of scientific discovery over decades. This progression has been shaped by various practical factors, including the expanding availability of chemical precursors, advances in synthesis and characterization methods, shifts in research paradigms, and evolving technological demands[22] (Fig. 1a).

Consequently, materials discovery reflects not only the intrinsic chemical feasibility of candidate materials but also the scientific and technological context of the time. Nevertheless, the chronology of materials discovery has rarely been utilized in materials informatics, with the time evolution of a materials stability network examined by Aykol et al.[18] as a notable exception, and its potential as a useful learning signal for prioritizing future materials discovery remains largely unexplored.

Here, we introduce ChronoSyn, a chronology-aware machine learning model for composition-based synthesizability prediction of hypothetical materials. ChronoSyn jointly learns to predict the synthesizability and, for experimentally reported materials, their relative reporting chronology through a shared composition representation. This joint objective lets the model learn patterns associated with the historical expansion of known materials space, providing additional source of supervision beyond conventional synthesizability labels (P or U).

We show that incorporating discovery chronology enables ChronoSyn to organize the learned materials space along the timeline of discovery, capturing distinct characteristics of earlier- and later-reported materials. This chronology-aware representation improves synthesizability prediction, and ablation studies confirm that the gains come from the incorporated chronological information. Moreover, when applied to a large hypothetical composition space (~50 million compositions), ChronoSyn identifies distinct chemical families with high predicted synthesizability. Follow-up literature analysis reveals that several high-scoring compositions within these families, although absent from the database and thus unseen during training, have in fact already been experimentally synthesized and reported. Together, these results demonstrate the potential of discovery chronology as an informative source of supervision for synthesizability prediction and highlight the practical applicability of ChronoSyn for prioritizing candidates for future discovery.

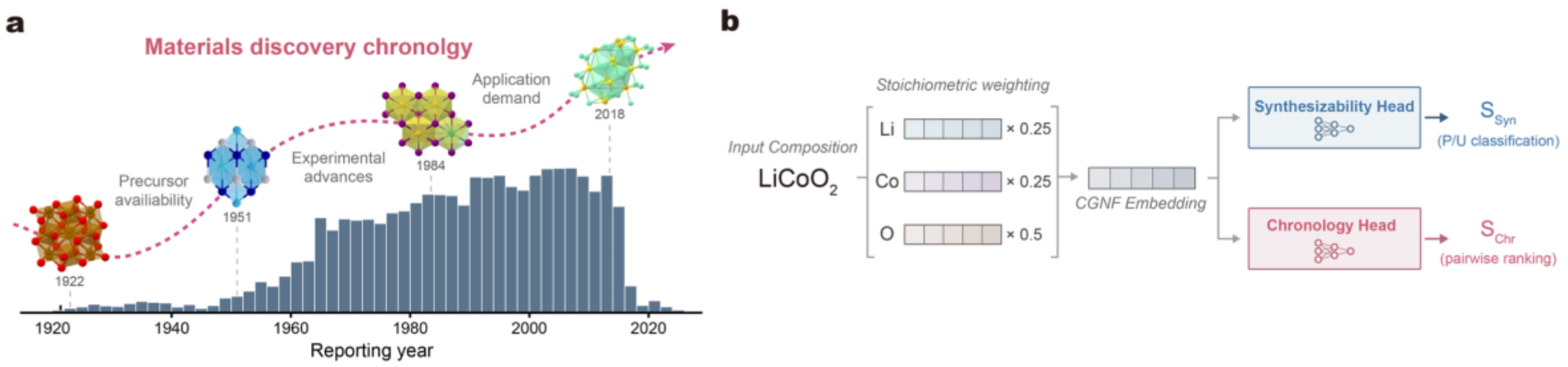

**Fig. 1. ChronoSyn overview. a.** Schematic illustration of the chronology underlying materials discovery. Materials discovery does not proceed through random exploration of chemical space, but follows trajectories shaped by practical factors such as precursor availability, advances in experimental methods, and application-driven demand. **b.** Architecture of ChronoSyn. A chemical composition is encoded into a CGNF embedding and passed through two heads for synthesizability prediction and pairwise chronological ranking.

# Results

## Chronology-aware synthesizability learning with ChronoSyn

The architecture of ChronoSyn largely follows the original model architecture reported by Jang et al.[17], with an additional chronology head introduced to learn materials discovery chronology (Fig. 1b). This design minimizes potential confounding effects arising from architectural modifications, allowing the contribution of the chronology signal to synthesizability prediction to be evaluated directly against the baseline model. Each input composition is represented as a Compositional Graph Neural Fingerprint (CGNF)[17,23], which computes a stoichiometry-weighted sum of element embeddings, and then passes it to two prediction heads. The synthesizability head outputs a synthesizability score ($S_{syn}$) used for the prediction of experimental synthesizability, while the chronology head outputs a chronology score ($S_{chr}$) for each material. The discovery chronology is learned using a pairwise-ranking objective, where, given a pair of materials, the model is trained to assign a higher $S_{chr}$ to the material reported more recently than to the one reported earlier.

To train ChronoSyn, we constructed datasets in which compositions were labeled as P or U according to whether they matched an experimentally reported entry in the Inorganic Crystal Structure Database (ICSD)[24]. For each P composition, the publication year was retrieved from the bibliographic metadata of the corresponding ICSD entry, which serves as a proxy for its discovery time. When a composition had been experimentally reported multiple times, the earliest publication year was assigned as its reporting year. Two datasets were constructed, differing in the source of the candidate composition space. The MP-only dataset draws compositions from the Materials Project (MP)[25], whereas the MP+OQMD dataset additionally incorporates compositions from the Open Quantum Materials Database (OQMD)[26,27], expanding the U pool more than sixfold. Using both datasets allowed us to examine whether the benefit of chronology generalizes beyond the composition space of a single database.

ChronoSyn was evaluated on each dataset against the corresponding ablation models (Table 1). The chronology-free baseline corresponds to the architecture of Jang et al.[17] without the chronology head. In two auxiliary-target ablation models, the chronology head is instead trained on randomly shuffled years or mean atomic mass information, respectively, representing either disruption of the original chronological signal or its replacement with a chemically meaningful but chronology-independent descriptor. Model performance was evaluated using the calibration procedure adopted in previous studies[16,17,28-30] to estimate classification metrics in the PU-learning setting, which depends on the class prior (α)[17] (see Methods). Across both datasets, ChronoSyn improved synthesizability prediction performance over the chronology-free baseline, indicating that incorporating chronology provides additional predictive information. Furthermore, comparisons with the auxiliary-target ablation models showed that this improvement was attributable not simply to the addition of an auxiliary head but to the actual discovery chronology signal. This advantage was robust to the choice of α, as ChronoSyn consistently outperformed all other models across a wide range of α values (Supplementary Fig. S4). Further details on the dataset construction, model training and evaluation metrics are provided in Methods.

Fig. 2a shows the prediction results of ChronoSyn (trained on the MP-only dataset) on the held-out test set. ChronoSyn assigned substantially higher synthesizability scores to experimentally reported (P) compositions than to unlabeled (U) compositions, resulting in a clear separation between the two groups. To examine whether the chronology objective is reflected in the learned representation, the latent embedding vectors of test P compositions from the chronology-free baseline and ChronoSyn were visualized using t-SNE[31] and colored according to their reporting years (Fig. 2b). To quantify how strongly the latent space is aligned with reporting chronology, we computed Spearman's rank correlation (ρ) between the latent embeddings and reporting years using ridge regression, which increased from 0.161 for the baseline to 0.505 for ChronoSyn (see Supplementary Section S3-1 for details). In the t-SNE projection of ChronoSyn, more recently reported compositions clustered in the lower-left region, whereas earlier-reported compositions clustered in the upper-right region. This confirms that the chronology signal is effectively encoded in the shared composition representation and generalizes to held-out compositions that were not seen during training. A similar trend was observed for the ChronoSyn model trained on the MP+OQMD dataset (Supplementary Fig. S6).

**Table 1. Synthesizability prediction performance of ChronoSyn and ablation models trained on the MP-only and MP+OQMD datasets.** The chronology-free baseline excludes the chronology head, while in the two auxiliary-target ablation models (year-shuffled and atomic-mass), the chronology head is instead trained on shuffled years and atomic mass information, respectively. Performance is reported as calibrated true positive rate (TPR) and precision on the held-out test set, with the decision threshold calibrated using a class prior of $\alpha = 0.081$ (see Methods). Under this calibration, TPR and precision are nearly identical, and a random classifier yields 0.081 for both. Results across α values are provided in Supplementary Fig. S4. The best value in each column is shown in bold. Given approximately 10,000 test P compositions in each dataset, the binomial standard error of each value is below 0.005, and the improvement in TPR of ChronoSyn over the chronology-free baseline is statistically significant in both datasets ($p < 10^{-46}$, McNemar's test[32]).

| Dataset | Model | TPR | Prec |
|---|---|---|---|
| MP-only | Baseline (Jang et al.[17]) | 0.470 | 0.470 |
| | ChronoSyn | **0.524** | **0.524** |
| | ChronoSyn (year-shuffled) | 0.484 | 0.484 |
| | ChronoSyn (atomic-mass) | 0.466 | 0.466 |
| MP+OQMD | Baseline (Jang et al.[17]) | 0.831 | 0.831 |
| | ChronoSyn | **0.873** | **0.873** |
| | ChronoSyn (year-shuffled) | 0.844 | 0.844 |
| | ChronoSyn (atomic-mass) | 0.838 | 0.838 |

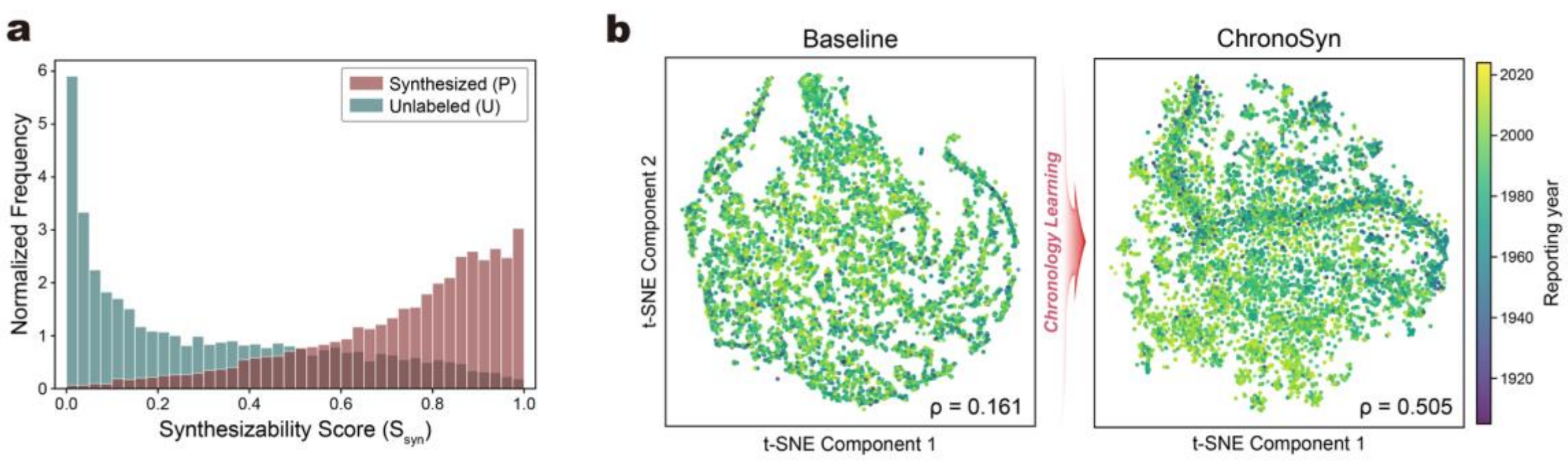

**Fig. 2. Synthesizability prediction and chronology-aware representation learning. a.** Distribution of synthesizability scores predicted by ChronoSyn (trained on the MP-only dataset) for experimentally reported (P) and unlabeled (U) compositions in the held-out test set. **b.** t-SNE visualization of the learned composition representations of test P compositions for the chronology-free baseline and ChronoSyn, colored by reporting year. Spearman's rank correlation (ρ) between the actual reporting years and those predicted from the full latent embeddings by cross-validated ridge regression is shown for each model.

## Impact of the learned chronology on U prediction

The preceding analysis of held-out test P compositions demonstrated that ChronoSyn learned a latent representation aligned with reporting chronology. We next examined how this learned discovery chronology relates to the predictions for U compositions, which lack explicit reporting year labels. Following the same approach as in Fig. 2b, ridge regression probes were fitted to map the latent embeddings of test P compositions to their reporting years and were then applied to the test U compositions to obtain a predicted reporting year for each composition (see Supplementary Section S3-1). The 10% of U compositions with the most recent predicted years were designated as recent-latent U, while the 10% with the earliest predicted years were designated as old-latent U (each 1,013 compositions). As shown in Fig. 3a, the two groups exhibited distinct spatial distributions in the t-SNE projection. Consistent with the pattern observed for test P compositions in Fig. 2b, recent-latent U compositions were predominantly distributed in the lower-left region, whereas old-latent U compositions were concentrated toward the upper-right region.

The predicted synthesizability scores of the recent-latent U and old-latent U compositions are shown in Fig. 3b. Recent-latent U compositions exhibited a substantially higher mean score than old-latent U compositions (0.34 vs 0.20), indicating that the learned chronology is associated with how ChronoSyn evaluates unreported compositions. This difference may reflect how thoroughly different chemical spaces have been explored. Chemical spaces resembling those of earlier-reported materials have been explored over a longer period, and compositions that remain unreported within them are more likely to be genuinely difficult to synthesize. By contrast, chemical spaces resembling those of recently reported materials are still under active exploration, and U compositions in these regions are more likely to include materials that have not yet been synthesized but could be experimentally realized.

The chemical characteristics of the compositions constituting the recent-latent and old-latent U groups were further compared, revealing distinct compositional profiles between the two groups (Fig. 3c). Recent-latent U compositions contained a greater number of elements on average than old-latent U compositions (4.20 vs 2.84) and more atoms per formula unit (16.5 vs 12.3). Meanwhile, they also exhibited a substantially lower mean atomic number (23.2 vs 40.9). These differences indicate that the two groups occupy chemically distinct regions of composition space, with recent-latent U compositions generally characterized by greater compositional complexity but lighter constituent elements than old-latent U compositions.

The two groups also exhibited pronounced differences in their most frequent elements (Fig. 3c). The most notable difference was the prevalence of oxygen. Only 19.1% of old-latent U compositions contained O, compared with 76.8% of recent-latent U compositions. Moreover, the elements most frequently found in recent-latent U compositions are closely associated with material families that have been actively studied in recent decades. Li, Fe, Mn and P are characteristic of phosphate-based cathode materials for lithium-ion batteries, which have been extensively investigated since the report of $LiFePO_4$ as a cathode material in 1997.[33] In parallel, Sr and Ca, together with Fe and Mn, are typical A- and B-site constituents of multicomponent perovskite oxides, which have been widely explored since the 2000s as electrode materials for solid oxide fuel cells[34] and as electrocatalysts for oxygen evolution[35]. The higher synthesizability scores assigned to recent-latent U compositions therefore suggest that ChronoSyn may implicitly capture signals associated with this evolving materials research frontier.

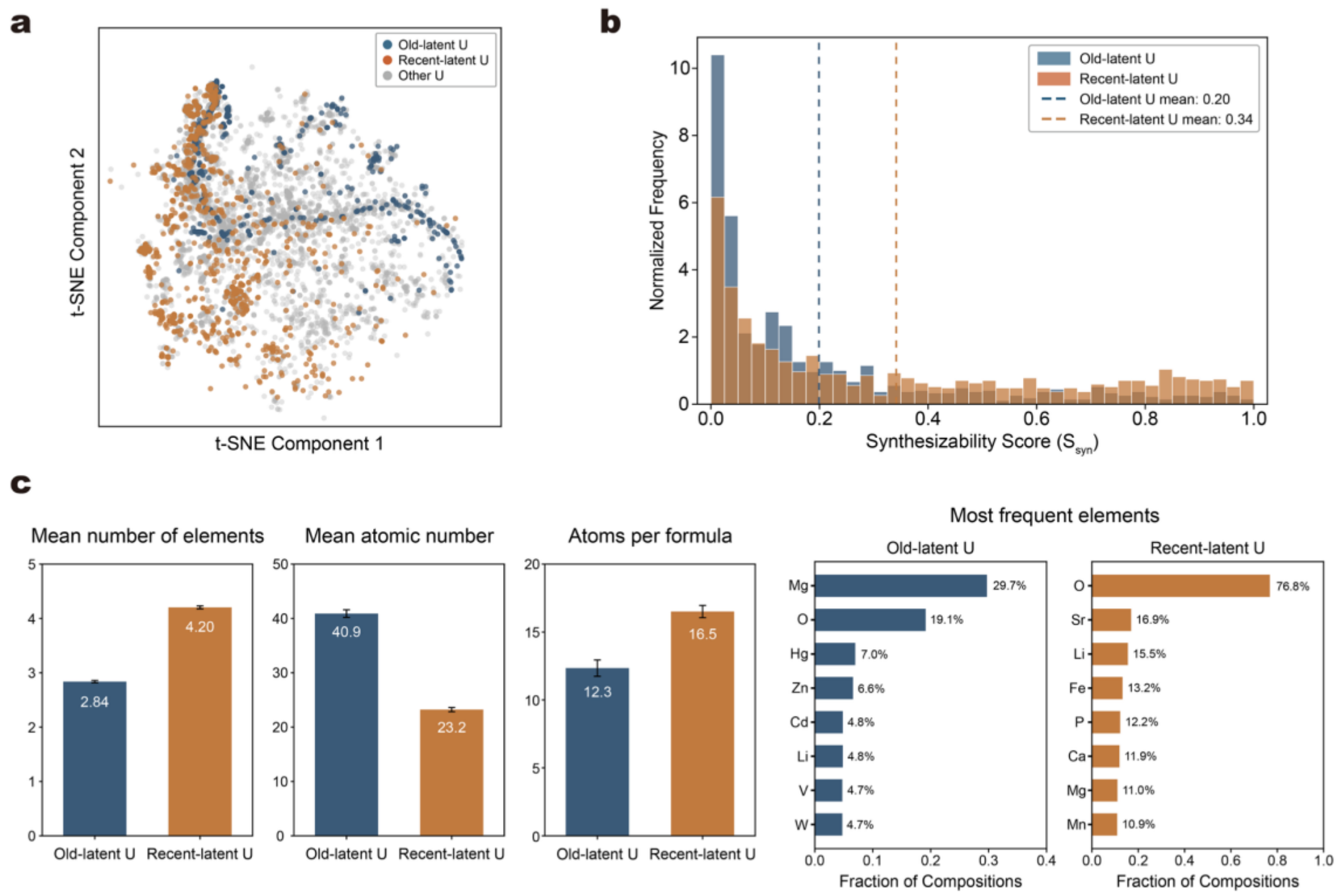

**Fig. 3. Applying learned discovery chronology to the unlabeled composition space. a.** t-SNE visualization of the held-out test unlabeled (U) compositions in the same projection as Fig. 2b. Old-latent and recent-latent U groups (1,013 compositions each) correspond to the 10% of U compositions with the earliest and most recent predicted reporting years, respectively, and the remaining U compositions are shown in gray. Predicted reporting years were obtained by applying ridge regression probes trained on test P compositions to the U compositions and averaging the predictions over all 100 bags (see Supplementary Section S3-1). **b.** Distribution of synthesizability scores for the old-latent and recent-latent U groups. Dashed lines indicate the mean score of each group. **c.** Compositional characteristics of the two groups, including the mean number of elements, mean atomic number, mean number of atoms per formula unit, and the most frequently occurring elements, shown as the fraction of compositions containing each element. Error bars denote the standard error of the mean (SEM).

## Chemical characteristics of compositions rescued by ChronoSyn

To identify the types of compositions most affected by the incorporation of chronology, we analyzed the test P compositions that were misclassified by the chronology-free baseline but correctly classified by ChronoSyn. The baseline model incorrectly classified 5,408 test P compositions as unsynthesizable (false negatives), of which 1,021 were correctly classified as synthesizable by ChronoSyn (Fig. 4a). (Conversely, 468 compositions correctly classified by the baseline were misclassified by ChronoSyn, yielding a net gain of 553 correctly identified P compositions.) To characterize which elemental classes were preferentially rescued, we compared the fraction of compositions containing at least one element of each class between the rescued and unrescued compositions (Fig. 4b). Compositions containing f-block elements showed the largest positive difference (15.5%), more than twice that of the next-ranked class, metalloids (7.1%). Further analysis of the learned representation revealed that f-block-containing compositions aligned more strongly with reporting chronology than other compositions (Supplementary Fig. S7), indicating that their reporting chronology is more strongly structured.

To understand the origin of this stronger chronological structure, we examined the reporting-year distributions of three representative f-block-containing material families (Fig. 4c). The first was permanent-magnet materials[36,37], represented by Sm–Co and Nd–Fe–B systems, in which rare-earth elements provide strong magnetocrystalline anisotropy. The second was lanthanide-containing cuprates (Ln–Cu–O, where Ln denotes lanthanide elements), including the well-known cuprate superconductors[38]. The third was colossal magnetoresistive (CMR) manganites[39] (Ln–Mn–O), in which lanthanide elements are combined with manganese-oxide frameworks. All three families showed pronounced temporal concentration in the materials database. For permanent-magnet compositions, the period from 1978 to 1992 accounted for 55.8% of all entries in this family. Cuprate compositions showed a concentrated reporting period from 1987 to 1996 (52.5%), whereas CMR manganites exhibited a later concentration between 1995 and 2008 (72.5%). These distributions reflect periods in which the exploration of each family expanded markedly. The preferential rescue of f-block-containing compositions therefore suggests that ChronoSyn may capture such family-specific discovery patterns, which are not represented by static P/U labels alone, enabling more accurate synthesizability predictions for compositions belonging to historically concentrated material families.

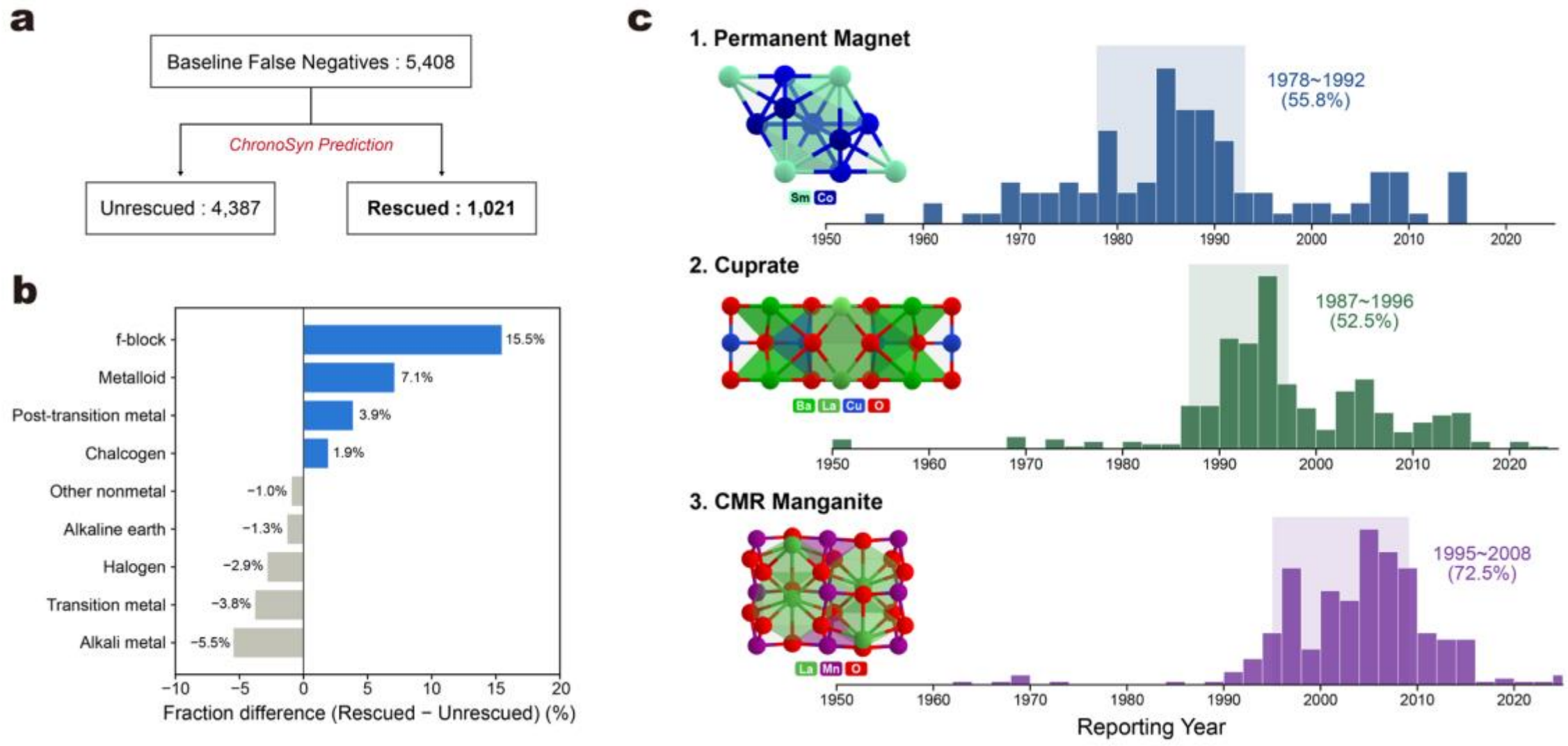

**Fig. 4. Chemical characteristics of compositions rescued by ChronoSyn. a.** Test P compositions incorrectly classified as unsynthesizable by the chronology-free baseline (false negatives) were re-evaluated by ChronoSyn, yielding 1,021 rescued and 4,387 unrescued compositions. **b.** Difference in the fraction of compositions containing at least one element of each elemental class between rescued and unrescued compositions. Elemental class definitions are listed in Supplementary Table S1. **c.** Reporting-year distributions of all database entries for representative f-block-containing material families, namely permanent magnets (Sm–Co and Nd–Fe–B), lanthanide-containing cuprates (Ln–Cu–O), where Ln denotes lanthanide elements, and colossal magnetoresistive (CMR) manganites (Ln–Mn–O), shown with representative crystal structures. Shaded regions indicate periods of concentrated reporting, with the fraction of entries reported within each period.

## Retrospective validation of materials discovery prioritization

To assess the practical utility of ChronoSyn in a setting that more closely resembles real-world materials discovery, we performed a retrospective validation. For a given cutoff year, the model was trained using only materials reported up to the cutoff, with both the synthesizability and chronology heads restricted to this period, and was then evaluated on its ability to identify compositions reported after the cutoff (Fig. 5a). This setup mirrors how researchers use the materials knowledge accumulated to date to anticipate and prioritize materials that may be realized in the future. Three cutoff years were considered (2015, 2018 and 2020). For each cutoff, the candidate pool consisted of P compositions reported after the cutoff together with a fixed set of held-out U compositions.

Performance was evaluated using Precision@*K*, defined as the fraction of the top-*K* ranked candidates that were reported after the cutoff, which reflects the hit rate under a fixed screening budget (Fig. 5b). ChronoSyn consistently achieved higher precision than the chronology-free baseline for all cutoffs and both values of *K*, and remained well above the level expected from random selection in all cases. The largest relative differences were observed among the highest-ranked candidates ($K = 100$) for the later cutoffs, albeit with greater statistical uncertainty. For the 2018 cutoff, for example, the baseline precision at $K = 100$ (0.03) was comparable to that of random selection (0.033), whereas ChronoSyn reached 0.07. These results suggest that the chronology learned from the historical materials database provides useful guidance for prioritizing compositions that had not yet been reported at the time of training, supporting the practical utility of chronology-aware synthesizability predictions.

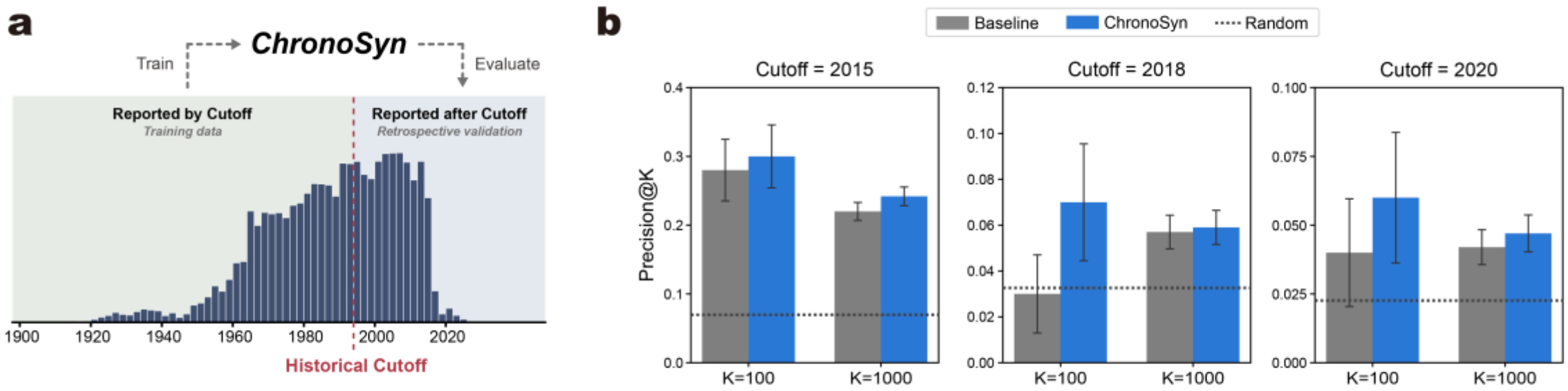

**Fig. 5. Retrospective validation of ChronoSyn for historical materials discovery. a.** Schematic of the historical time-split evaluation, overlaid on the reporting-year distribution of experimentally reported compositions. At each cutoff year, ChronoSyn is trained using materials reported up to the cutoff and evaluated on its ability to prioritize compositions reported after the cutoff. **b.** Precision@$K$ of the chronology-free baseline and ChronoSyn at historical cutoffs of 2015, 2018 and 2020 for $K$ = 100 and 1000. The candidate pool for each cutoff comprises P compositions reported after the cutoff ($n$ = 807, 363 and 249, respectively) and a fixed set of held-out U compositions ($n$ = 10,765). Error bars denote the standard error of the binomial proportion. Dashed lines indicate the expected precision of random selection.

## ChronoSyn-guided exploration of the vast hypothetical composition space

Next, the ability of ChronoSyn to directly explore a vast hypothetical composition space and identify regions with high predicted synthesizability was evaluated. Following the framework of Park et al.[40], the hypothetical composition space was constructed by combinatorial enumeration of the 89 elements used in ChronoSyn training. Each composition was categorized into four groups (Standard, Interesting, Missing, and Unlikely) according to its presence in the MP database and its validity under the SMACT[9,41] charge-neutrality criterion (Fig. 6a). This procedure resulted in a large hypothetical composition space comprising approximately 50 million binary and ternary compositions (Table 2).

The Missing and Unlikely compositions were subsequently evaluated using ChronoSyn trained on the MP-only dataset. Because neither group contains compositions present in MP, both provide an independent composition space for evaluating the model's predictions. As shown in Fig. 6b, ChronoSyn assigned noticeably higher synthesizability scores to Missing compositions than to Unlikely compositions, with median scores of 0.539 vs 0.435 (binary) and 0.552 vs 0.401 (ternary). This difference is consistent with the charge-neutrality criterion that distinguishes the two groups, suggesting that ChronoSyn has implicitly captured information related to charge neutrality from the training data, even though no such constraint was explicitly imposed.

We next analyzed the Missing compositions in greater detail. Fig. 6c shows a t-SNE projection of the latent embeddings of the binary Missing compositions, colored according to their predicted synthesizability scores, in which high-scoring compositions are concentrated in several regions. To delineate these visually apparent regions, K-means clustering was applied, and the chemical families associated with high-scoring clusters were identified. Two representative regions are highlighted in Fig. 6c. Region 1 mainly comprised lanthanide or actinide elements combined with late transition metals, frequently in a 1:2 ratio. This pattern is characteristic of Laves-phase intermetallics[42], whose formation is favored by an atomic radius ratio of approximately 1.225 between the larger and smaller atoms (consistent with the combination of large lanthanide/actinide elements with smaller late transition metals). Region 2 was dominated by f-block metals combined with platinum-group metals, intermetallic systems whose formation can be rationalized by the large negative formation enthalpies predicted by the Miedema model[43], arising largely from the pronounced electronegativity contrast between the two groups of metals. Details of the clustering analysis and descriptions of the remaining high-scoring clusters are provided in Supplementary Section S4-1.

Notably, a literature survey of selected high-scoring compositions revealed that several compositions within these high-scoring regions had already been experimentally reported despite being absent from MP. For example, the $YbAu_2$ composition in Region 1 was reported in 1966 as part of a systematic X-ray survey of rare-earth-containing binary alloys[44], and URu in Region 2 was reported in 1968 in a phase diagram study of uranium–platinum-group systems[45]. These examples illustrate that the high-scoring regions identified by ChronoSyn can contain experimentally realizable compositions, highlighting its potential as a guide for exploring the vast hypothetical composition space.

**Table 2. Summary of the hypothetical composition space.** Binary and ternary compositions were classified into four categories following Park et al.[40], according to their validity under the SMACT charge-neutrality criterion and their presence in the Materials Project (MP). Standard and Interesting compositions are SMACT-valid and SMACT-invalid compositions present in MP, respectively, whereas Missing and Unlikely compositions are SMACT-valid and SMACT-invalid compositions absent from MP, respectively. Counts and percentages (relative to the total number of binary or ternary compositions) are reported for each category, together with the total number of unique compositions.

| System | Metric | SMACT: Valid | | SMACT: Invalid | | Total Unique |
|---|---|---|---|---|---|---|
| | | **MP: Present** | **MP: Absent** | **MP: Present** | **MP: Absent** | |
| Binary ($A_xB_y$) | Status | Standard | Missing | Interesting | Unlikely | 168,388 |
| | Count (%) | 3,680 (2.19%) | 7,485 (4.45%) | 6,510 (3.87%) | 150,713 (89.50%) | |
| Ternary ($A_xB_yC_z$) | Status | Standard | Missing | Interesting | Unlikely | 49,854,596 |
| | Count (%) | 25,040 (0.05%) | 8,037,364 (16.12%) | 12,574 (0.03%) | 41,779,618 (83.80%) | |

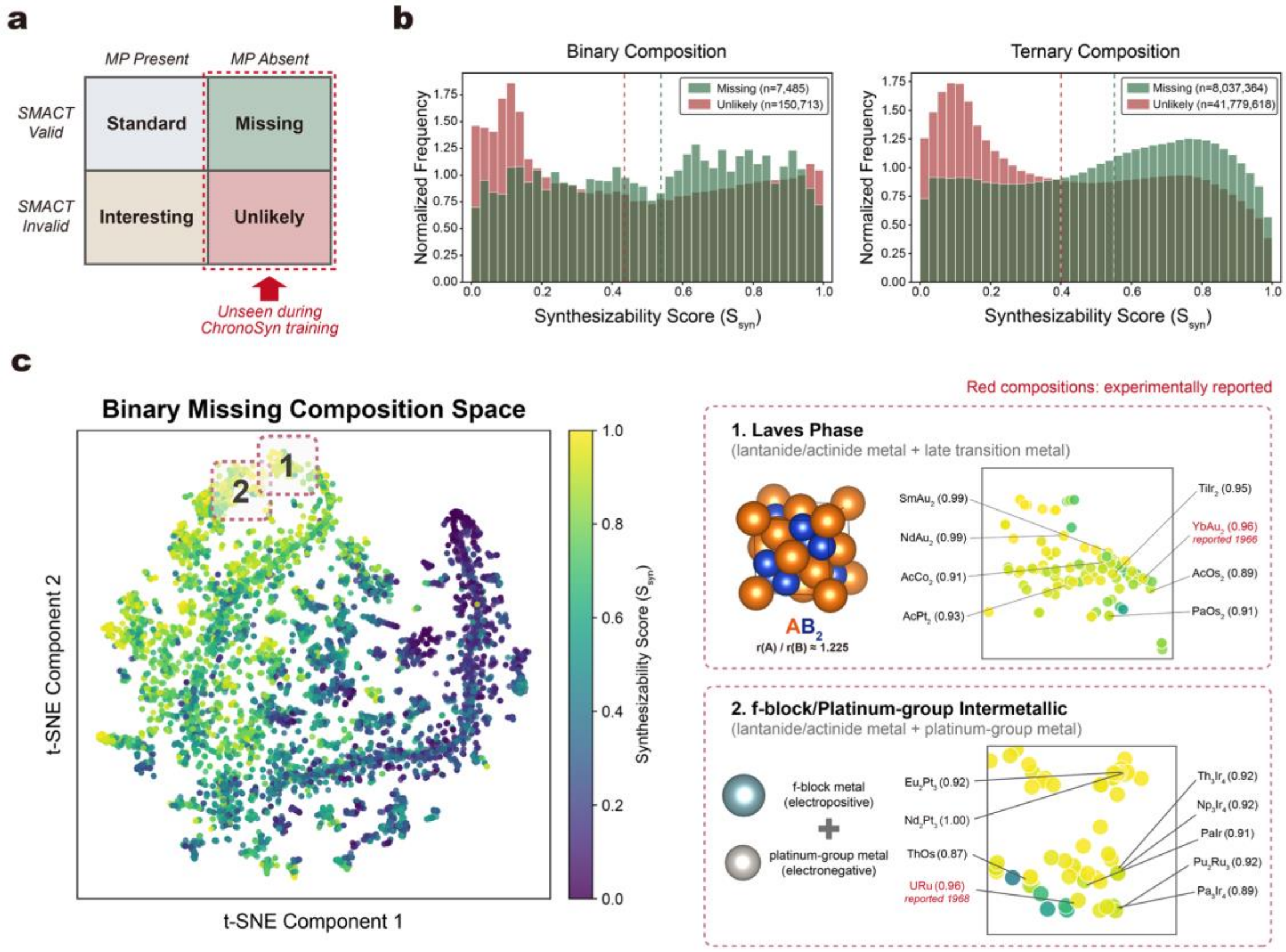

**Fig. 6. ChronoSyn screening of the composition space absent from the Materials Project. a.** Classification of enumerated compositions according to their presence in the Materials Project (MP) and validity under the SMACT charge-neutrality criterion. Missing and Unlikely compositions were not included in ChronoSyn training. **b.** Distributions of ChronoSyn synthesizability scores for Missing and Unlikely compositions in the binary and ternary composition spaces. Dashed lines indicate the median score of each group. **c.** t-SNE visualization of the binary Missing composition space, colored by synthesizability score. Representative high-scoring regions, delineated by K-means clustering of the t-SNE coordinates, are highlighted together with their characteristic chemical families and example compositions. Red labels indicate compositions absent from MP but experimentally reported in the literature, with their reporting years.

A similar analysis was subsequently performed for the ternary Missing compositions. Given their much larger number (~8 million), applying K-means clustering as in the binary analysis was computationally impractical. Instead, the most frequently occurring elemental pairs in the top-scoring compositions were first identified, and their associated third elements were examined to generalize these combinations into broader chemical families. Four representative families among the ten highest-scoring ones are highlighted in Fig. 7 as Regions 1–4. Details of the analysis and descriptions of the remaining top-scoring families are provided in Supplementary Section S4-2.

Region 1 contained Na–S compositions combined with Si, Ge, or P, a chemical space extensively explored for Na-ion solid electrolytes.[46] It included recently reported compounds absent from MP with distinct structural motifs, such as the argyrodite-type framework ($Na_7PS_6$[47]) and the isolated-tetrahedral compounds ($Na_4SiS_4$[48] and $Na_4GeS_4$[49]). Region 2 consisted of polar intermetallics combining rare-earth metals and Mg with Sn, Bi, or Sb, in which the electropositive metals transfer valence electrons to the more electronegative p-block elements. Reported examples included $Eu_2Mg_3Bi_4$[50] and GdMgSn[51], studied for their magnetic properties, and LaMgSn[52], investigated for hydrogen storage. Region 3 contained compositions of rare-earth metals with C and N in stoichiometries consistent with carbodiimide chemistry, eight of which were reported by Neukirch et al.[53] in 2006 through a systematic solid-state metathesis study. Region 4 comprised alkali or alkaline-earth compositions with P and N in stoichiometries consistent with nitridophosphate chemistry, which is typically built from $PN_4$ tetrahedra. Among the reported examples, $CaP_2N_4$[54] was reported as a host material for $Eu^{2+}$ luminescence, and $Sr_3P_3N_7$[55] was reported in a methodological study comparing ammonothermal and high-pressure synthesis routes.

Beyond these individual examples, 101 binary and 1,990 ternary Missing compositions were found to have experimental reports in ICSD despite being absent from MP, providing an independent reference set for evaluating exploration performance. ChronoSyn correctly identified more of these compositions as synthesizable than the chronology-free baseline, with increases of 5.9 and 3.6 percentage points for binary and ternary compositions, respectively (Supplementary Section S4-3), suggesting that the benefit of chronology also extend to the hypothetical composition space beyond existing databases. Taken together, these results illustrate that ChronoSyn captures chemically meaningful synthesizability patterns across diverse chemistries and prioritizes compositions that are likely to be realizable.

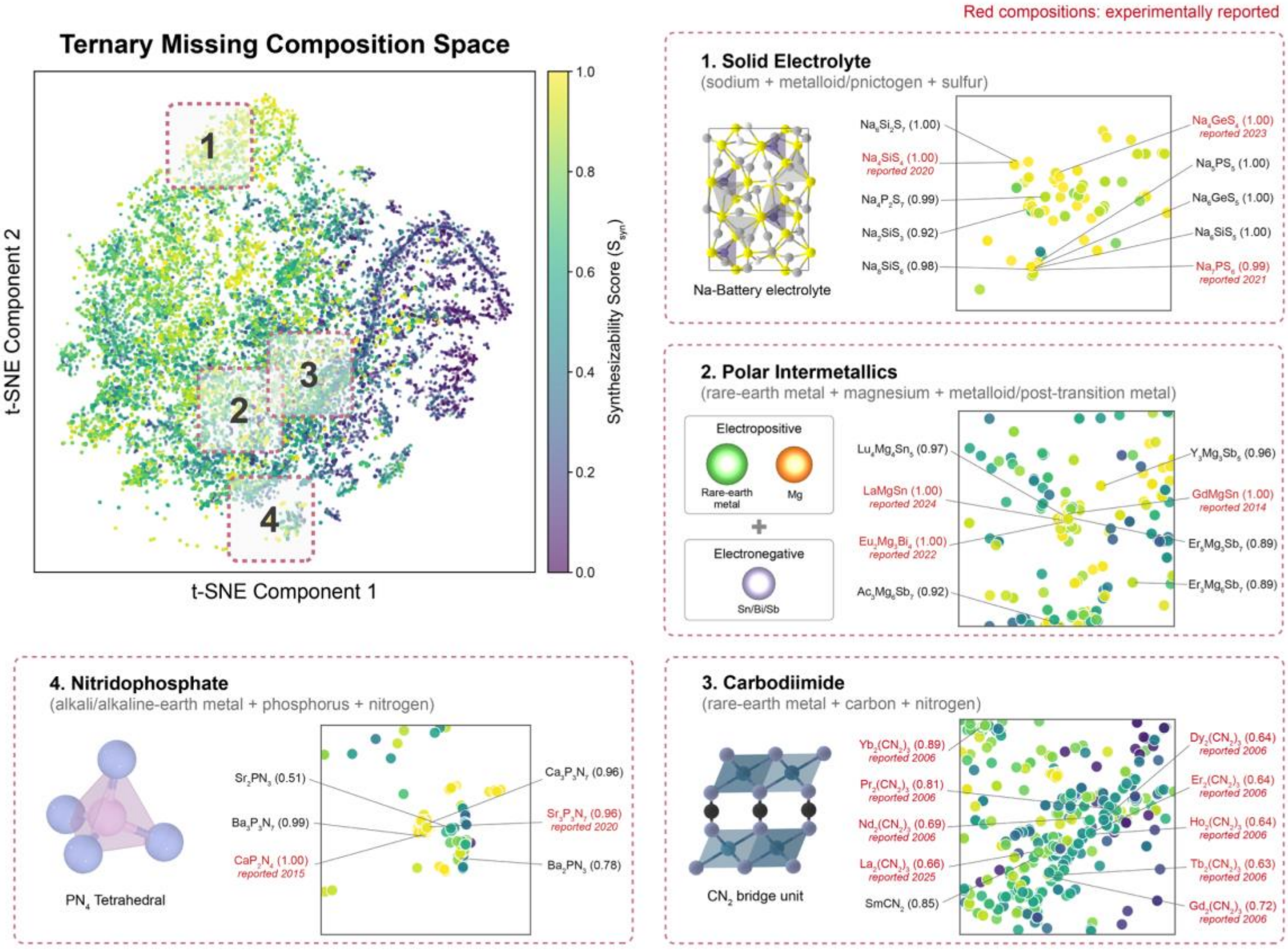

**Fig. 7. ChronoSyn screening of the ternary Missing composition space.** t-SNE visualization of a random subset of 20,000 ternary Missing compositions, together with all compositions belonging to the four highlighted chemical families, colored by ChronoSyn synthesizability score. Four representative high-scoring regions, corresponding to chemical families identified from frequently occurring elemental combinations among top-scoring compositions, are highlighted together with their characteristic chemical families and example compositions, with synthesizability scores shown in parentheses. Red labels indicate compositions absent from MP but experimentally reported in the literature, with their reporting years.

## Discussion

We have found that the chronology of materials discovery provides an informative source of supervision beyond conventional positive-unlabeled synthesizability labels. By learning how chemical-space exploration has unfolded over time, ChronoSyn evaluates each composition not in isolation but within the historical trajectory of discovery, distinguishing well-explored regions from emerging ones and capturing the temporally concentrated discovery patterns in material families. In this sense, the history of materials discovery serves not merely as a record of what has been found, but as a guide to where future discoveries are likely to emerge.

Several limitations define the scope of the present approach. First, the reporting year used here is an observable proxy for discovery chronology and may not fully reflect when a material was first synthesized or recognized. Second, although the chronology signal captures practical factors that shape experimental realization, it may also lead ChronoSyn to favor chemical spaces that have historically attracted research attention, potentially assigning lower scores to chemically viable compositions in regions that have received little attention to date. This may limit its ability to identify entirely new classes of materials outside established research directions. Finally, ChronoSyn operates at the composition level and does not explicitly account for structural polymorphism, which can strongly influence whether a given material is experimentally realizable or if exploration into a novel family of structure types has been achieved.

Despite these limitations, the chronology objective could be readily incorporated into structure-based synthesizability models[13] as well as other data-driven approaches. Recent synthesizability prediction models based on large language models (LLMs)[16,28-30] are particularly promising candidates, as they can draw on broad chemical knowledge acquired during pre-training. Although ChronoSyn achieved higher precision in an approximate comparison with such a model (Supplementary Section S2-2), the two approaches are complementary rather than competing. More broadly, our results suggest that the temporal dimension of materials data, typically discarded in current materials informatics, encodes how different regions of chemical space have been opened, expanded, and matured over time. This perspective may also extend to other fields with rich historical records of discovery, such as drug discovery and catalysis, where the trajectory of past discoveries may hold clues to what lies ahead.

# Methods

## Dataset construction

We used three databases to construct the training data. Compositions were obtained from the Materials Project (MP)[25], queried in June 2026, and from the Open Quantum Materials Database (OQMD)[26,27], using the composition pool of Jang et al.[17] Experimental reports were obtained from the Inorganic Crystal Structure Database (ICSD)[24], snapshot from July 2025 excluding theoretical structures. We constructed two datasets (MP-only and MP+OQMD), which differ in the databases used to define the candidate composition space. For each dataset, unique compositions were obtained by deduplicating reduced formulas and matching them against experimental ICSD entries to assign P/U labels. The MP-only dataset restricts both P and U to unique reduced compositions present in MP, yielding 104,837 compositions (51,010 P and 53,827 U). The MP+OQMD dataset additionally incorporates unique compositions from OQMD into the candidate composition space, expanding the dataset to 408,638 compositions (51,651 P and 356,987 U). The number of P compositions remains comparable between the two datasets, because only a small fraction of the additional OQMD compositions (641 of 303,801) had an ICSD match. The compositional characteristics and element coverage of the two datasets are shown in Supplementary Figs. S1 and S2.

For each P composition, the publication year was retrieved from the bibliographic metadata of the corresponding ICSD entries. When a composition corresponded to multiple ICSD entries, the earliest publication year was assigned as its reporting year. The resulting reporting years range from 1905 to 2024 (Supplementary Fig. S3). Two limitations of this procedure should be noted. First, because the MP snapshot is more recent than the ICSD release, compositions present in MP whose experimental reports were added to ICSD after July 2025 may have been labeled as U. Such cases are consistent with the PU-learning formulation, in which unlabeled data are assumed to contain undiscovered positives, and would, if anything, lead to a conservative estimate of model performance. Second, as shown in Figs. 1a and 5a and Supplementary Fig. S3, the number of P compositions drops sharply in the 2020s rather than continuing the increasing trend of earlier decades. We attribute this to the coverage limits of MP and ICSD, compounded by the fact that the original MP structures were largely sourced from ICSD. This under-representation of recent discoveries also limits the data available for learning the most recent part of the discovery chronology, and incorporating more up-to-date database snapshots in future work could mitigate this censoring effect.

## Model architecture and training

ChronoSyn, as illustrated in Fig. 1b, extends the composition-based PU-learning baseline of Jang et al.[17] with an additional chronology head attached to the same shared trunk. The trunk maps the 90-dimensional CGNF[17,23] composition feature into a 450-dimensional hidden representation through five fully connected layers with Softplus activation, consisting of an input projection layer followed by four 450-dimensional hidden layers. This shared hidden representation is passed to two heads, namely a binary classification head, trained with a negative log-likelihood loss to distinguish P from U, and an auxiliary chronology head that outputs a scalar chronology score ($S_{chr}$) for each composition.

The chronology head was trained with a pairwise ranking loss computed only over P compositions, since U compositions have no associated reporting year. Let $\mathcal{P}$ denote the set of P-composition pairs $(i, j)$ for which $y_i > y_j$, where $y$ is the reporting year. The loss penalizes cases in which the predicted chronology scores violate a margin $m$,

$$\mathrm{L_{chr}} = \frac{1}{|\mathcal{P}|} \sum_{(\mathrm{i,j}) \in \mathcal{P}} \max\left(0, \mathrm{m} - \left(\mathrm{S_{chr,i}} - \mathrm{S_{chr,j}}\right)\right)$$

where $m = 0.1$. To decouple this ranking signal from the small, noisy subset of P samples that happen to co-occur within any single classification mini-batch, a fresh set of 512 P compositions was drawn with replacement from the training P set at every training step, independently of the classification mini-batch, and $\mathcal{P}$ was constructed from all pairs within this set with different reporting years (up to approximately 130,000 pairs). The total training objective is

$$L_{total} = L_{PU} + \lambda L_{chr}$$

where $\lambda = 3.0$ was used throughout this work.

Ablation models were trained with identical architectures and hyperparameters. The chronology-free baseline omits the chronology head, which is equivalent to setting $\lambda = 0$. In the year-shuffled model, reporting years were randomly permuted among training P compositions, and in the atomic-mass model, the chronology head was trained to rank compositions by mean atomic mass instead of reporting year.

For both the MP-only and MP+OQMD datasets, we used a stratified 80/20 train/test split (stratified by P/U label). In place of validation-based model selection, we followed the PU-bagging scheme of Jang et al.[17,56] and trained a 100-model bootstrap-aggregating ensemble. For each of 100 bags, a bootstrap subset of training U equal in size to the training P set was resampled with replacement, and a freshly initialized model was trained on the resulting bag for a fixed 100 epochs using the Adam optimizer[57] with a learning rate of $10^{-3}$ and a batch size of 128. The final prediction for each composition was the average over all 100 bags.

## Performance evaluation

Under the PU setting, standard precision cannot be computed directly from held-out test data, since U is a mixture of true negatives and as-yet-unconfirmed positives rather than a clean negative class. We therefore estimated precision using the correction of Jain et al.[58], which recovers an estimate of true precision from quantities observable in PU data, given the class prior α (the true fraction of positives within U). For a given threshold $t$, $TPR^{PU}$ was defined as the fraction of held-out P scored above $t$ and $FPR^{PU}$ as the fraction of held-out U scored above $t$. True precision is then estimated as

$$\text{Prec} = \alpha \cdot TPR^{PU} / FPR^{PU}$$

We set $\alpha = 0.081$, as estimated by Jang et al.[17] for the MP+OQMD composition space, and applied the same value to the MP-only dataset following previous studies[30]. Because MP-only U compositions are likely enriched in undiscovered positives relative to the much larger OQMD-derived pool, this choice is expected to yield a conservative estimate of precision on the MP-only dataset.

The decision threshold for each model was calibrated so that the top α fraction of held-out U falls above it, that is, at the $(1 - \alpha)$ quantile of the U score distribution. This ensures that all models are compared at the same rate of positive predictions on U. Under this calibration, $FPR^{PU} \approx \alpha$, so that the estimated precision is nearly identical to $TPR^{PU}$, and a random classifier yields $TPR^{PU} \approx \text{Prec} \approx \alpha$. To assess the sensitivity of the results to the choice of α, the calibration was repeated over a range of α values, and ChronoSyn outperformed all other models across this range (Supplementary Fig. S4).

### Construction of the hypothetical composition space

The hypothetical composition space analyzed in Figs. 6 and 7 was constructed following the framework of Park et al.[40] Binary and ternary compositions were exhaustively enumerated over the 89 elements used in ChronoSyn training, with stoichiometric coefficients of up to 8 per element, yielding 168,388 binary and 49,854,596 ternary compositions. Each composition was then evaluated using SMACT[9,41] (v.4.0.0), which tests charge neutrality and electronegativity balance based on its default common oxidation states, and was classified as SMACT-valid or SMACT-invalid. Compositions were further classified according to whether their reduced formula was present in MP, yielding four groups. SMACT-valid and SMACT-invalid compositions present in MP were labeled Standard and Interesting, respectively, whereas SMACT-valid and SMACT-invalid compositions absent from MP were labeled Missing and Unlikely, respectively. The Missing and Unlikely compositions, none of which were included in ChronoSyn training, were scored using the ChronoSyn ensemble trained on the MP-only dataset. Details of the clustering and chemical family analyses are provided in Supplementary Sections S4-1 and S4-2.

## Data availability

The datasets used to train the model in this study are publicly available in the Github repository at https://github.com/parkjunkil/ChronoSyn.

## Code availability

The source code and trained models developed in this study are publicly available in the Github repository at https://github.com/parkjunkil/ChronoSyn.

# Reference

1 Merchant, A. *et al.* Scaling deep learning for materials discovery. *Nature* **624**, 80–85 (2023).
2 Schmidt, J. *et al.* Machine-learning-assisted determination of the global zero-temperature phase diagram of materials. *Advanced Materials* **35**, 2210788 (2023).
3 Barros-Luque, L. *et al.* The Open Materials 2024 (OMat24) inorganic materials dataset and models. *Nature Computational Science*, 1–11 (2026).
4 Zeni, C. *et al.* A generative model for inorganic materials design. *Nature* **639**, 624–632 (2025).
5 Xie, T., Fu, X., Ganea, O.-E., Barzilay, R. & Jaakkola, T. Crystal diffusion variational autoencoder for periodic material generation. *arXiv preprint arXiv:2110.06197* (2021).
6 Jiao, R. *et al.* Crystal structure prediction by joint equivariant diffusion. *Advances in Neural Information Processing Systems* **36**, 17464–17497 (2023).
7 Park, J., Lee, Y. & Kim, J. Multi-modal conditional diffusion model using signed distance functions for metal-organic frameworks generation. *Nature Communications* **16**, 34 (2025).
8 Park, J., Choi, J. & Jung, Y. Generative modelling of inorganic materials with explicit electronic structure. *Nature Communications* (2026).
9 Davies, D. W. *et al.* Computational screening of all stoichiometric inorganic materials. *Chem* **1**, 617–627 (2016).
10 George, J. *et al.* The limited predictive power of the Pauling rules. *Angewandte Chemie* **132**, 7639–7645 (2020).
11 Sun, W. *et al.* The thermodynamic scale of inorganic crystalline metastability. *Science advances* **2**, e1600225 (2016).
12 Aykol, M., Dwaraknath, S. S., Sun, W. & Persson, K. A. Thermodynamic limit for synthesis of metastable inorganic materials. *Science advances* **4**, eaaq0148 (2018).
13 Jang, J., Gu, G. H., Noh, J., Kim, J. & Jung, Y. Structure-based synthesizability prediction of crystals using partially supervised learning. *Journal of the American Chemical Society* **142**, 18836–18843 (2020).
14 Frey, N. C. *et al.* Prediction of synthesis of 2D metal carbides and nitrides (MXenes) and their precursors with positive and unlabeled machine learning. *ACS nano* **13**, 3031–3041 (2019).
15 Antoniuk, E. R. *et al.* Predicting the synthesizability of crystalline inorganic materials from the data of known material compositions. *npj Computational Materials* **9**, 155 (2023).
16 Kim, S., Jung, Y. & Schrier, J. Large language models for inorganic synthesis predictions. *Journal of the American Chemical Society* **146**, 19654–19659 (2024).
17 Jang, J. *et al.* Synthesizability of materials stoichiometry using semi-supervised learning. *Matter* **7**, 2294–2312 (2024).
18 Aykol, M. *et al.* Network analysis of synthesizable materials discovery. *Nature communications* **10**, 2018 (2019).
19 Elkan, C. & Noto, K. in *Proceedings of the 14th ACM SIGKDD international conference on Knowledge discovery and data mining.* 213–220.
20 Bekker, J. & Davis, J. Learning from positive and unlabeled data: A survey. *Machine learning* **109**, 719–760 (2020).
21 Jia, X. *et al.* Anthropogenic biases in chemical reaction data hinder exploratory inorganic synthesis. *Nature* **573**, 251–255 (2019).
22 Agrawal, A. & Choudhary, A. Perspective: Materials informatics and big data: Realization of the "fourth paradigm" of science in materials science. *APL materials* **4** (2016).
23 Goodall, R. E. & Lee, A. A. Predicting materials properties without crystal structure: deep representation learning from stoichiometry. *Nature communications* **11**, 6280 (2020).
24 Belsky, A., Hellenbrandt, M., Karen, V. L. & Luksch, P. New developments in the Inorganic Crystal Structure Database (ICSD): accessibility in support of materials research and design. *Structural Science* **58**, 364–369 (2002).
25 Jain, A. *et al.* Commentary: The Materials Project: A materials genome approach to accelerating materials innovation. *APL materials* **1** (2013).
26 Saal, J. E., Kirklin, S., Aykol, M., Meredig, B. & Wolverton, C. Materials design and discovery with high-throughput density functional theory: the open quantum materials database (OQMD). *Jom* **65**, 1501–1509 (2013).
27 Kirklin, S. *et al.* The Open Quantum Materials Database (OQMD): assessing the accuracy of DFT formation energies. *npj Computational Materials* **1**, 15010 (2015).
28 Kim, S., Schrier, J. & Jung, Y. Explainable synthesizability prediction of inorganic crystal polymorphs using large language models. *Angewandte Chemie International Edition* **64**, e202423950 (2025).
29 Choi, J., Kim, S. & Jung, Y. Synthesis-aware materials redesign via large language models. *Journal of the American Chemical Society* **147**, 39113–39122 (2025).
30 Choi, J. *et al.* Compiling Chemical Knowledge into Executable Descriptors for Materials Prediction. *arXiv preprint arXiv:2608.27587* (2026).
31 Van der Maaten, L. & Hinton, G. Visualizing data using t-SNE. *Journal of machine learning research* **9** (2008).
32 McNemar, Q. Note on the sampling error of the difference between correlated proportions or percentages. *Psychometrika* **12**, 153–157 (1947).
33 Padhi, A. K., Nanjundaswamy, K. S. & Goodenough, J. B. Phospho-olivines as positive-electrode materials for rechargeable lithium batteries. *Journal of the electrochemical society* **144**, 1188–1194 (1997).
34 Shao, Z. & Haile, S. M. A high-performance cathode for the next generation of solid-oxide fuel cells. *Nature* **431**, 170–173 (2004).
35 Suntivich, J., May, K. J., Gasteiger, H. A., Goodenough, J. B. & Shao-Horn, Y. A perovskite oxide optimized for oxygen evolution catalysis from molecular orbital principles. *Science* **334**, 1383–1385 (2011).
36 Strnat, K., Hoffer, G., Olson, J., Ostertag, W. & Becker, J. A family of new cobalt-base permanent magnet materials. *Journal of Applied Physics* **38**, 1001–1002 (1967).
37 Sagawa, M., Fujimura, S., Togawa, N., Yamamoto, H. & Matsuura, Y. New material for permanent magnets on a base of Nd and Fe. *Journal of applied physics* **55**, 2083–2087 (1984).
38 Bednorz, J. G. & Müller, K. A. Possible high T c superconductivity in the Ba− La− Cu− O system. *Zeitschrift für physik B condensed matter* **64**, 189–193 (1986).
39 Jin, S. *et al.* Thousandfold change in resistivity in magnetoresistive La-Ca-Mn-O films. *Science* **264**, 413–415 (1994).
40 Park, H., Onwuli, A., Butler, K. T. & Walsh, A. Mapping inorganic crystal chemical space. *Faraday Discussions* **256**, 601–613 (2025).
41 Davies, D. W. *et al.* Smact: Semiconducting materials by analogy and chemical theory. *Journal of Open Source Software* **4**, 1361 (2019).
42 Stein, F., Palm, M. & Sauthoff, G. Structure and stability of Laves phases. Part I. Critical assessment of factors controlling Laves phase stability. *Intermetallics* **12**, 713–720 (2004).
43 Miedema, A. R., de Châtel, P. F. & de Boer, F. R. Cohesion in alloys—fundamentals of a semi-empirical model. *Physica B+ c* **100**, 1–28 (1980).
44 Moriarty, J., Humphreys, J., Gordon, R. & Baenziger, N. X-ray examination of some rare-earth-containing binary alloy systems. *Acta Crystallographica* **21**, 840–841 (1966).
45 Park, J. J. Reactions of Uranium With the Platinide Elements. I. The Uranium-Ruthenium System. *Journal of Research of the National Bureau of Standards. Section A, Physics and Chemistry* **72**, 1 (1968).
46 Hayashi, A., Noi, K., Sakuda, A. & Tatsumisago, M. Superionic glass-ceramic electrolytes for room-temperature rechargeable sodium batteries. *Nature communications* **3**, 856 (2012).
47 Studenyak, I. *et al.* Structural, electrical and optical properties of ion-conducting Na6PS5Cl, Na6PS5Br, and Na7PS6 compounds. *Journal of Physics and Chemistry of Solids* **159**, 110269 (2021).
48 Harm, S. *et al.* Finding the right blend: interplay between structure and sodium ion conductivity in the system Na5AlS4–Na4SiS4. *Frontiers in Chemistry* **8**, 90 (2020).
49 Yahia, H. B., Motohashi, K., Mori, S., Sakuda, A. & Hayashi, A. Synthesis, structure and properties of Na4GeS4. *Journal of Alloys and Compounds* **960**, 170600 (2023).
50 Marshall, M. *et al.* Eu2Mg3Bi4: Competing Magnetic Orders on a Buckled Honeycomb Lattice. *Chemistry of Materials* **34**, 3902–3909 (2022).
51 Lemoine, P. *et al.* Neutron powder diffraction and Mössbauer spectroscopy (119 Sn and 155 Gd) studies of the CeScSi-type GdMgSn and GdMgPb compounds. *Physical Review B* **89**, 174428 (2014).
52 Yartys, V. A. *et al.* Structure and bonding in TiNiSi type LaMgSnH intermetallic hydride. *Journal of Alloys and Compounds* **976**, 173198 (2024).
53 Neukirch, M., Tragl, S. & Meyer, H. J. Syntheses and Structural Properties of Rare Earth Carbodiimides. *Inorganic Chemistry* **45**, 8188–8193 (2006). https://doi.org/10.1021/ic0608952
54 Pucher, F. J., Marchuk, A., Schmidt, P. J., Wiechert, D. & Schnick, W. Luminescent Nitridophosphates CaP2N4: Eu2+, SrP2N4: Eu2+, BaP2N4: Eu2+, and BaSr2P6N12: Eu2+. *Chemistry–A European Journal* **21**, 6443–6448 (2015).
55 Mallmann, M., Wendl, S., Strobel, P., Schmidt, P. J. & Schnick, W. Sr3P3N7: Complementary Approach by Ammonothermal and High-Pressure Syntheses. *Chemistry–A European Journal* **26**,

6257–6263 (2020).
56 Mordelet, F. & Vert, J.-P. A bagging SVM to learn from positive and unlabeled examples. *Pattern recognition letters* **37**, 201–209 (2014).
57 Kingma, D. P. & Ba, J. Adam: A method for stochastic optimization. *arXiv preprint arXiv:1412.6980* (2014).
58 Jain, S., White, M. & Radivojac, P. in *Proceedings of the AAAI Conference on Artificial Intelligence.*

# Learning the Chronology of Materials Discovery for Synthesizability Prediction

Junkil Park[1*] and Aron Walsh[1*]

[1] Department of Materials, Imperial College London, London SW7 2AZ, U.K.

**Supplementary Information**

# Section S1. Details on Dataset Construction

## Section S1-1. Compositional Characteristics of the Datasets

The two datasets used in this study were constructed from the Materials Project (MP)[1], the Open Quantum Materials Database (OQMD)[2,3] and the Inorganic Crystal Structure Database (ICSD)[4], as described in Methods. Supplementary Fig. S1 shows the distribution of compositions by the number of constituent elements in the two datasets. In both datasets, ternary compositions were the most common, followed by quaternary compositions, with the fraction of ternary compositions being considerably higher in the MP+OQMD dataset (72.6%) than in the MP-only dataset (45.2%). This difference reflects the composition of the additional OQMD-derived pool, which consists predominantly of hypothetical ternary compositions and therefore shifts the overall distribution of the MP+OQMD dataset toward ternary systems.

Supplementary Fig. S2 shows the fraction of compositions containing each element. Both datasets cover the same set of 89 elements used in ChronoSyn training, while elements not included in this set, mainly radioactive heavy elements such as Po, At, Rn, Fr, Ra and the actinides beyond Pu, are shown in gray. In the MP-only dataset, O was by far the most frequent element, appearing in 47% of compositions, consistent with the prevalence of oxides among experimentally studied inorganic materials.[4] In the MP+OQMD dataset, the fraction of O-containing compositions was considerably lower (15%), and the elemental distribution was more uniform across the periodic table, reflecting the broad combinatorial coverage of the OQMD-derived hypothetical compositions.

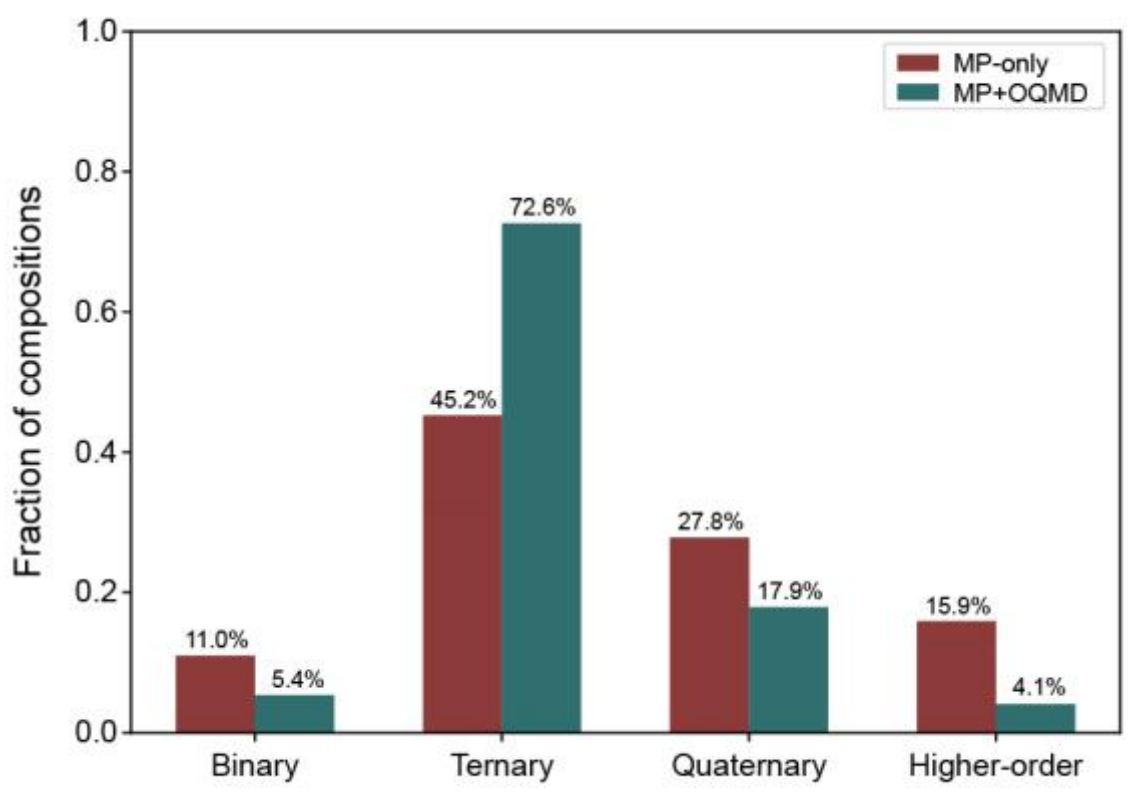

**Supplementary Fig. S1. Distributions of compositions by number of constituent elements.** Fraction of binary, ternary, quaternary and higher-order compositions in the MP-only ($n$ = 104,837) and MP+OQMD ($n$ = 408,638) datasets. Higher-order compositions denote those containing five or more elements. Fractions are calculated relative to the total number of unique compositions in each dataset.

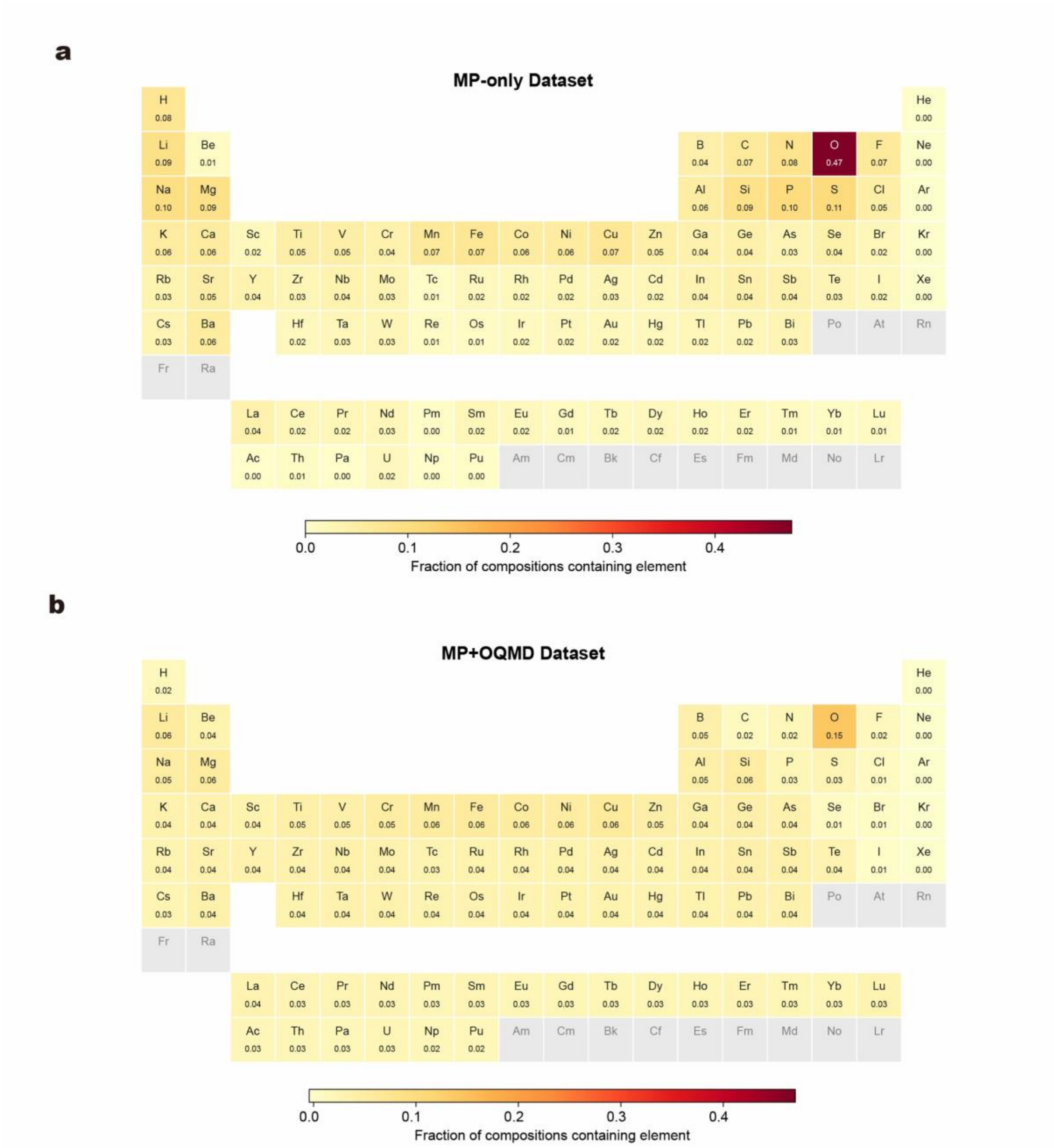

**Supplementary Fig. S2. Element coverage of the training datasets.** Fraction of compositions containing each element in **a**, the MP-only dataset and **b**, the MP+OQMD dataset. Values in each cell indicate the fraction of compositions in which the corresponding element appears. Gray cells denote elements not included among the 89 elements used in ChronoSyn training.

## Section S1-2. Reporting-Year Distributions of the Datasets

Supplementary Fig. S3 shows the distribution of reporting years for the P compositions in the two datasets. Because the two datasets share most of their P compositions (51,010 and 51,651, respectively), their distributions are nearly identical. Few compositions were reported before the 1950s, after which the number of reported compositions increased rapidly, reaching a maximum in the 2000s and early 2010s. In contrast, the number of compositions reported in the 2020s is markedly lower than in preceding decades. As discussed in Methods, this is attributed to the coverage limits of MP and ICSD rather than to a genuine slowdown in experimental discovery.

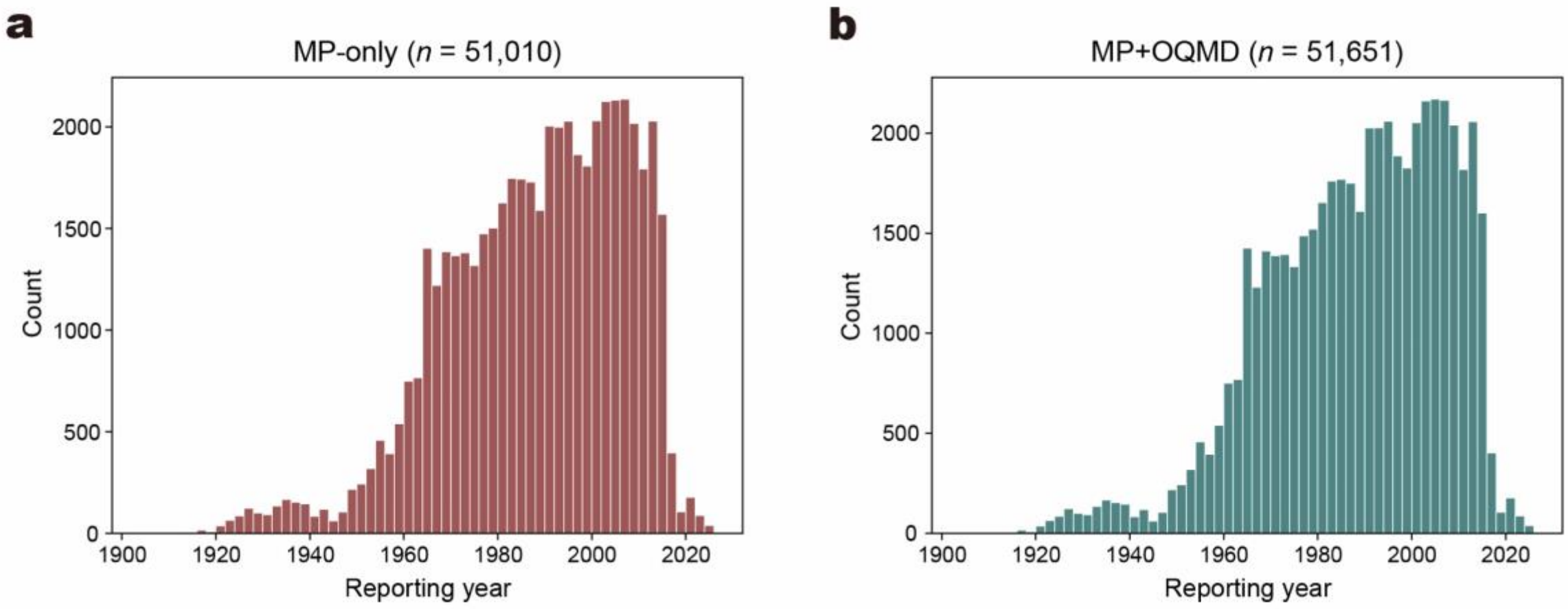

**Supplementary Fig. S3. Reporting-year distributions of P compositions.** Number of P compositions as a function of reporting year in **a**, the MP-only dataset and **b**, the MP+OQMD dataset. The reporting year of each composition corresponds to the earliest publication year among its matching ICSD entries. Histograms are shown in 2-year bins.

# Section S2. Details on Model Evaluation

## Section S2-1. Sensitivity to the Class Prior

The calibrated performance reported in Table 1 depends on the class prior α, which determines the decision threshold for each model[5] (see Methods). Because α = 0.081 was estimated by Jang et al.[6] for the MP+OQMD composition space and was also applied to the MP-only dataset, we examined whether the relative performance of the models depends on this choice. The calibration was repeated for α values ranging from 0.01 to 0.5, and true positive rate (TPR) was computed for each model at each value of α (Supplementary Fig. S4). Under this calibration, α corresponds to the fraction of U compositions predicted as positive, so that the curves trace how TPR changes as the screening criterion is relaxed. In both datasets, ChronoSyn achieved the highest TPR across the full range of α examined. The differences between ChronoSyn and the other models were most evident at small α, corresponding to the most stringent selection of top-ranked candidates. These results indicate that the advantage of ChronoSyn reported in Table 1 is not specific to the chosen value of α.

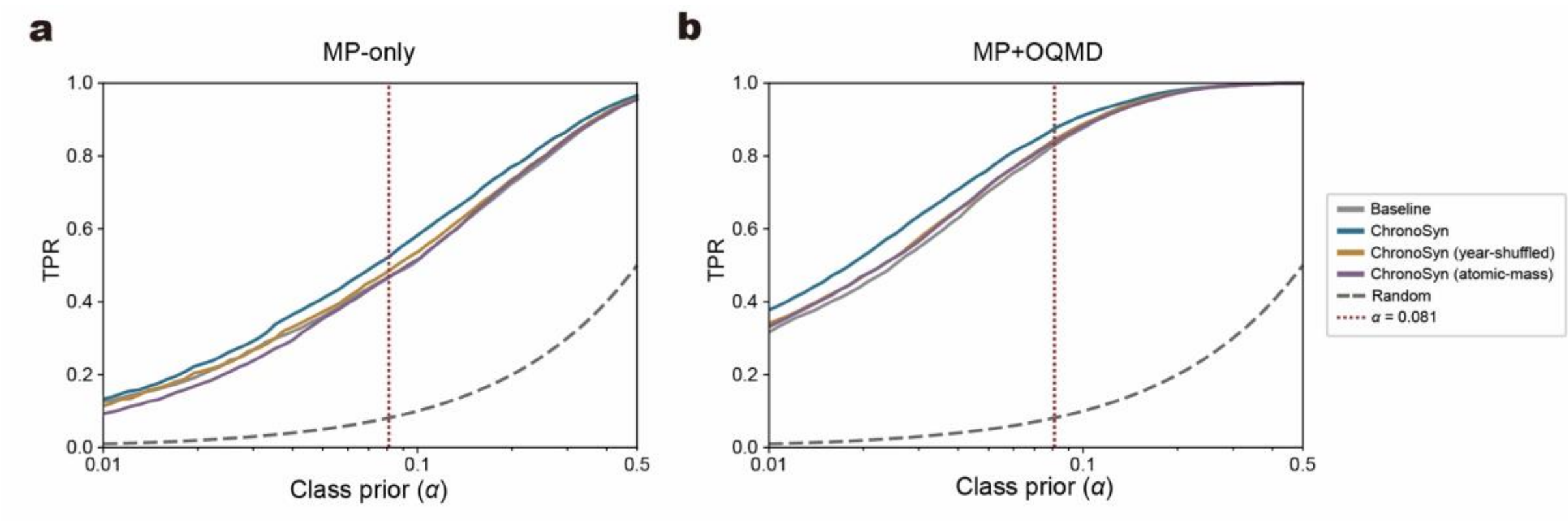

**Supplementary Fig. S4. Sensitivity of model performance to the class prior.** TPR of the chronology-free baseline, ChronoSyn and the two auxiliary-target ablation models, labeled ChronoSyn (year-shuffled) and ChronoSyn (atomic-mass), as a function of the class prior α used for the threshold calibration, for **a**, the MP-only dataset and **b**, the MP+OQMD dataset. Under this calibration, α corresponds to the fraction of held-out U compositions predicted as positive, and the estimated precision is nearly identical to TPR. The gray dashed line indicates the expected TPR of a random classifier (TPR = α), and the red dotted line indicates α = 0.081, the value used in this study. The x axis is shown on a logarithmic scale.

## Section S2-2. Comparison with an LLM-Based Approach

To benchmark ChronoSyn against a recent approach based on large language models (LLMs), we compared ChronoSyn trained on the MP+OQMD dataset with HEM-LLM reported by Yoon et al.[7] To our knowledge, HEM-LLM is the only existing LLM-based synthesizability model that takes chemical compositions as input. HEM-LLM formulates synthesizability prediction as a binary text-classification task, in which a fine-tuned LLM (Qwen3-14B) is prompted to classify a given composition as synthesizable (P) or non-synthesizable (N), and a continuous score is obtained from the difference between the log-sum-exp of the logits assigned to the "P" and "N" tokens. As in ChronoSyn, its positive labels are defined by the presence of an ICSD reference, whereas its negative labels are compositions assigned low scores by a previously published synthesizability model[6], rather than unlabeled compositions.

Because the two models were trained on largely overlapping but non-identical data, the comparison was restricted to compositions unseen by both models. From the held-out test set of ChronoSyn trained on the MP+OQMD dataset, we retained only compositions whose reduced formulas do not appear in the training or validation data of HEM-LLM, yielding 9,376 compositions (1,245 P and 8,131 U). Both models were evaluated against the same labels, in which U compositions were treated as negatives. Because the two models produce scores on different scales, performance was compared using Precision@$K$, which depends only on the ranking of candidates and not on how each model's scores are calibrated.

This comparison has several limitations. Although both models were trained on compositions drawn largely from MP and OQMD, they differ in their learning formulation (PU learning versus binary classification with curated negatives). In addition, because LLMs are pretrained on large text corpora, HEM-LLM may have been exposed to literature reporting some of the compositions in the comparison set, which would favor the LLM-based approach. The results should therefore be regarded as an approximate comparison rather than a controlled benchmark. Within these limitations, ChronoSyn achieved higher Precision@$K$ than HEM-LLM at all values of $K$ examined (Supplementary Fig. S5), with both models well above the precision expected from random selection (0.133).

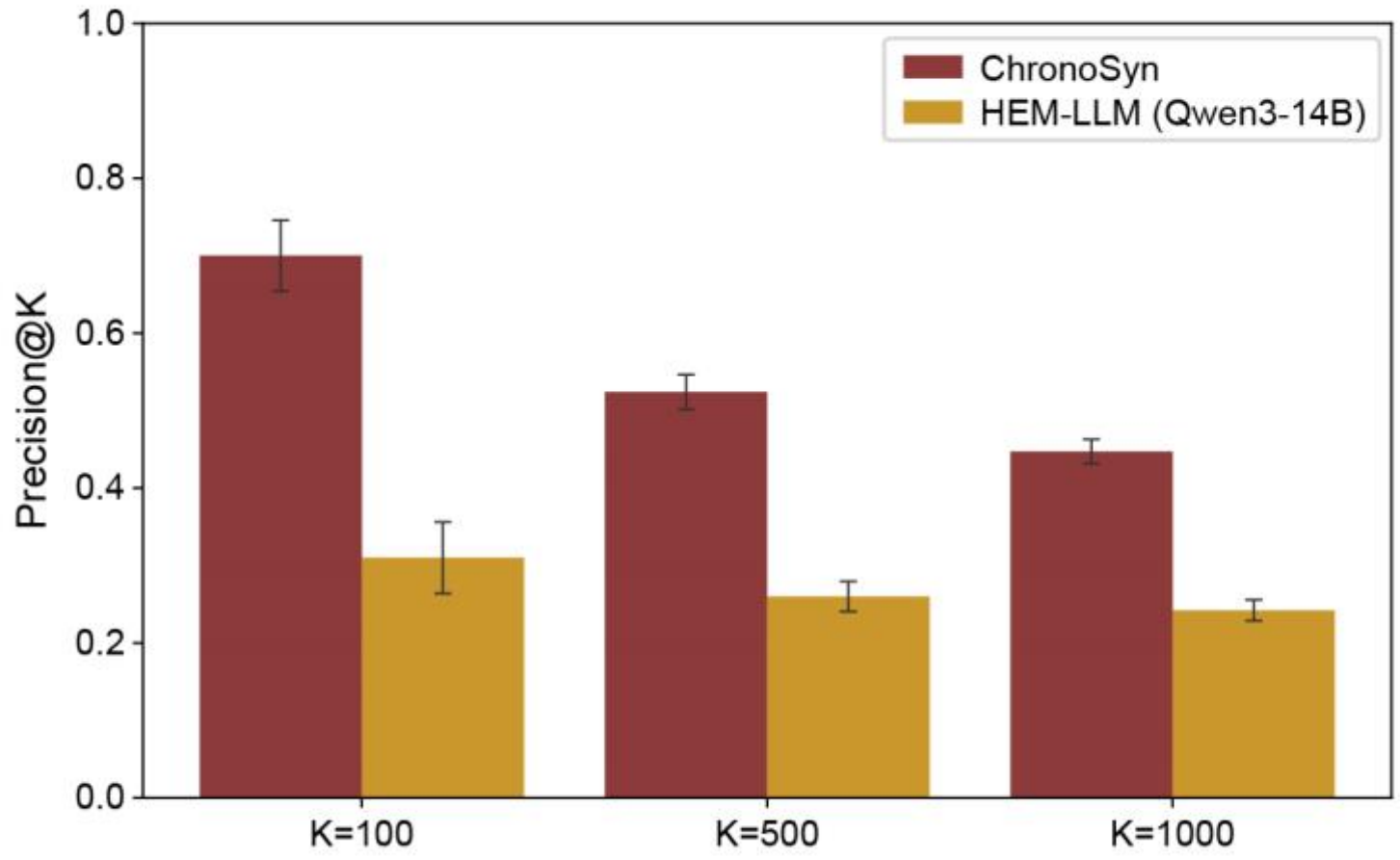

**Supplementary Fig. S5. Comparison of ChronoSyn with an LLM-based synthesizability prediction model.** Precision@*K* of ChronoSyn (trained on the MP+OQMD dataset) and HEM-LLM (Qwen3-14B) for *K* = 100, 500 and 1000, evaluated on 9,376 compositions (1,245 P and 8,131 U) unseen by both models during training. Error bars denote the standard error of the binomial proportion.

# Section S3. Details on Chronology-Aware Representation Analysis

## Section S3-1. Ridge Probe and t-SNE Settings

To quantify the alignment between the learned representation and reporting chronology, ridge regression models (scikit-learn[8], regularization strength of 10) were fitted to predict the normalized reporting year from the 450-dimensional hidden representation of test P compositions. Because each independently trained bag learns its own representation basis, representations from different bags cannot be meaningfully combined, and probes were therefore fitted separately for each bag. For the correlation analysis in Fig. 2b, a probe was fitted to the representation of a single bag using 5-fold cross-validation, and Spearman's ρ was computed on the out-of-fold predictions. For the analysis of U compositions in Fig. 3, a probe was fitted to all test P compositions for each of the 100 bags and applied to the corresponding representations of test U compositions, and the predicted reporting year of each U composition was obtained by averaging the predictions over all bags. For visualization, the hidden representations of the same bag were embedded into two dimensions by t-SNE[9] (openTSNE[10], perplexity of 30) after reduction to 50 dimensions by PCA, and test U compositions in Fig. 3 were projected onto the same embedding.

## Section S3-2. Representation Analysis of the MP+OQMD Model

To examine whether the trends observed for the MP-only dataset (Fig. 2) generalize to a larger composition space, the same analyses were performed for the models trained on the MP+OQMD dataset (Supplementary Fig. S6). As in the MP-only case, ChronoSyn assigned substantially higher synthesizability scores to P compositions than to U compositions, resulting in a clear separation between the two groups (Supplementary Fig. S6a). The alignment between the learned representation and reporting chronology also increased markedly upon incorporating the chronology objective, with Spearman's ρ rising from 0.241 for the baseline to 0.558 for ChronoSyn (Supplementary Fig. S6b), consistent with the increase observed for the MP-only dataset (0.161 to 0.505). These results confirm that the chronology signal is effectively encoded in the shared composition representation regardless of the composition space used for training.

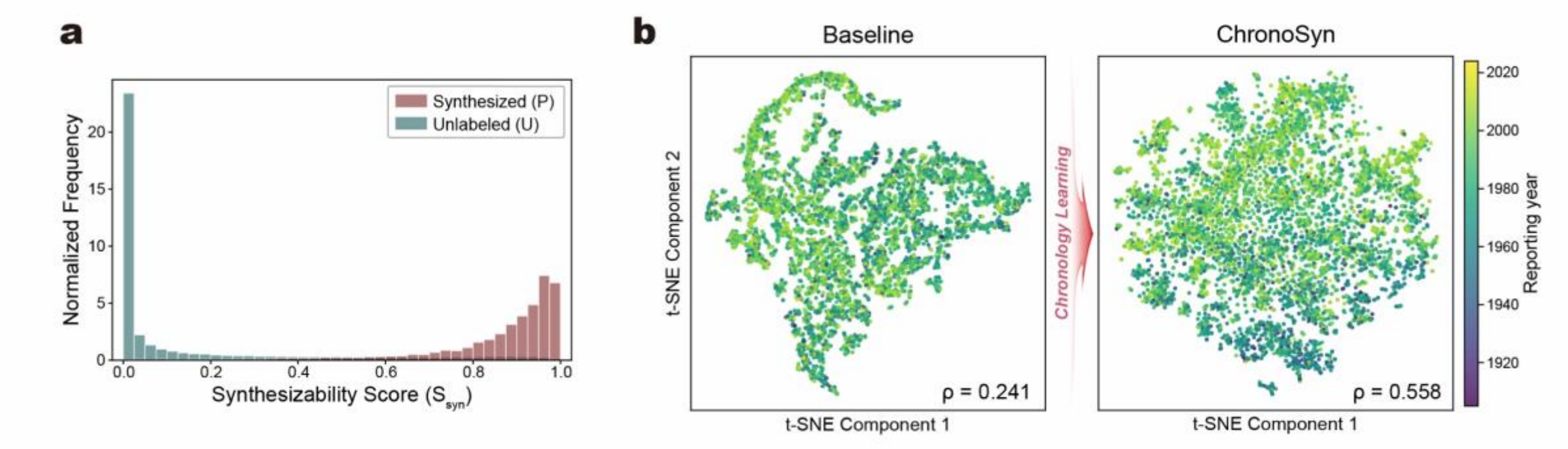

**Supplementary Fig. S6. Synthesizability prediction and chronology-aware representation learning on the MP+OQMD dataset. a.** Distribution of synthesizability scores predicted by ChronoSyn (trained on the MP+OQMD dataset) for experimentally reported (P) and unlabeled (U) compositions in the held-out test set. **b.** t-SNE visualization of the learned composition representations of test P compositions for the chronology-free baseline and ChronoSyn, colored by reporting year. Spearman's rank correlation (ρ) between the actual reporting years and those predicted from the full latent embeddings by cross-validated ridge regression is shown for each model.

## Section S3-3. Details on the Rescued-Sample Analysis

For the rescued-sample analysis in Fig. 4b, each of the 89 elements used in ChronoSyn training was assigned to a single elemental class, as listed in Supplementary Table S1. Because the boundaries between some classes vary across the literature, the following conventions were adopted. La and Ac were included in the f-block, and Zn, Cd and Hg were classified as transition metals. Te, which is sometimes regarded as a metalloid, was assigned only to the chalcogens, so that no element belonged to more than one class. For each class, the fraction of compositions containing at least one element of that class was computed separately for the rescued and unrescued compositions, and the difference between the two fractions (rescued − unrescued) is shown in Fig. 4b. Noble gases were excluded from Fig. 4b because compositions containing them were extremely rare.

To further examine the preferential rescue of f-block-containing compositions, we compared the alignment between the learned representation of ChronoSyn and reporting chronology separately for test P compositions with and without f-block elements (Supplementary Fig. S7). For each group, a ridge regression probe was fitted independently using 5-fold cross-validation, following the procedure described in Supplementary Section S3-1, and Spearman's $\rho$ was computed on the out-of-fold predictions. Compositions with f-block elements showed stronger alignment with reporting chronology ($\rho = 0.599$) than those without f-block elements ($\rho = 0.446$), despite the smaller number of compositions available for fitting the probe. These results suggest that the reporting chronology of f-block-containing compositions is more strongly structured, consistent with their preferential rescue by ChronoSyn.

**Supplementary Table S1. Elemental class assignments used in the rescued-sample analysis (Fig. 4b).** Each of the 89 elements used in ChronoSyn training was assigned to a single class.

| Category | Members |
|---|---|
| f-block | La–Lu, Ac–Pu |
| Metalloid | B, Si, Ge, As, Sb |
| Alkaline earth | Be, Mg, Ca, Sr, Ba |
| Alkali metal | Li, Na, K, Rb, Cs |
| Post-transition metal | Al, Ga, In, Sn, Tl, Pb, Bi |
| Transition metal | Sc–Zn, Y–Cd, Hf–Hg |
| Chalcogen | O, S, Se, Te |
| Halogen | F, Cl, Br, I |
| Other nonmetal | H, C, N, P |
| Noble gas | He, Ne, Ar, Kr, Xe |

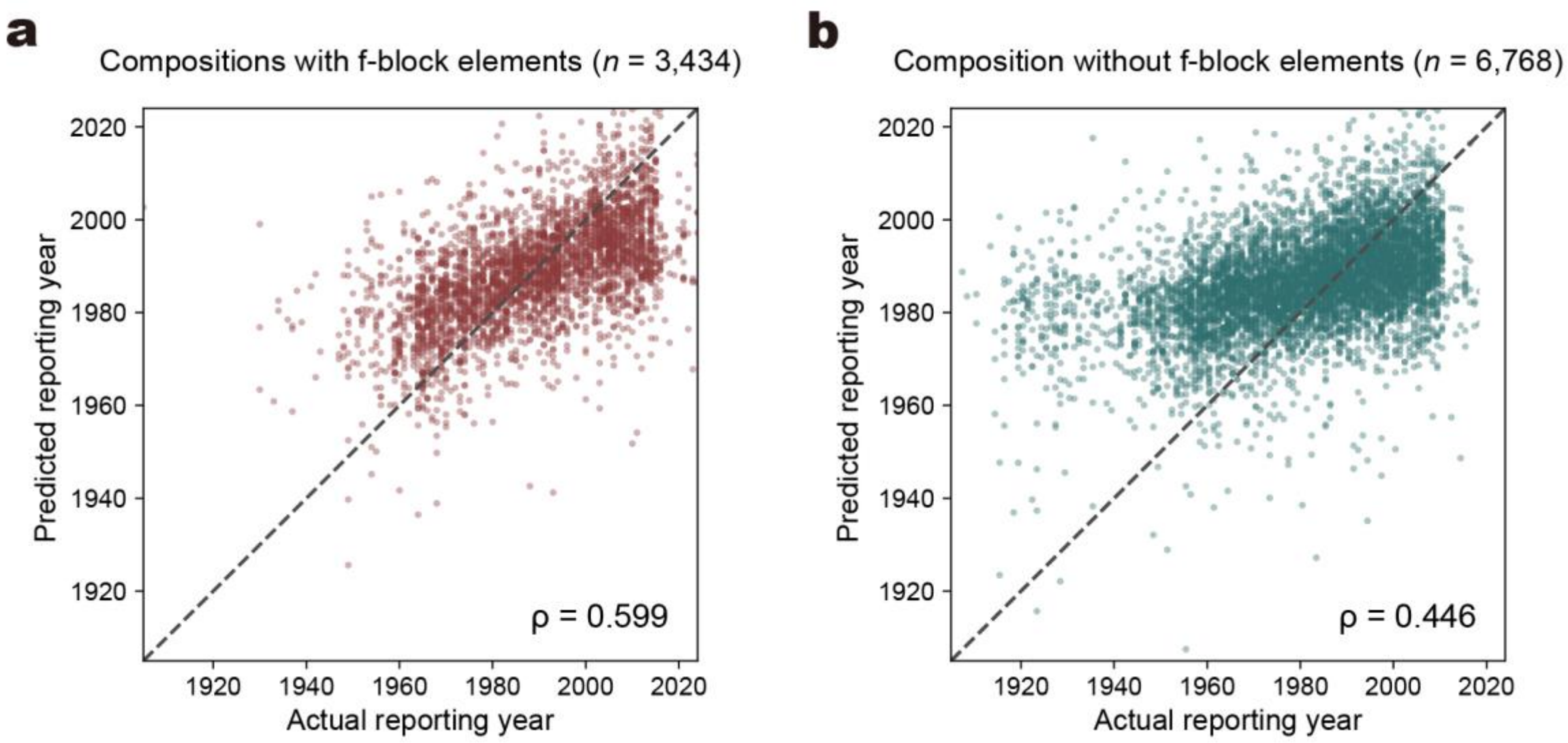

**Supplementary Fig. S7. Alignment between the learned representation and reporting chronology for f-block-containing and other compositions.** Reporting years predicted from the learned representations of test P compositions in ChronoSyn trained on the MP-only dataset versus actual reporting years, for **a**, compositions with f-block elements and **b**, compositions without f-block elements. For each group, a ridge regression probe was fitted independently, and predictions were obtained by 5-fold cross-validation. Spearman's rank correlation (ρ) between predicted and actual reporting years is shown in each panel. Dashed lines indicate perfect agreement.

# Section S4. Details on Hypothetical Composition Space Exploration

## Section S4-1. Clustering Analysis of Binary Missing Compositions

To identify regions of high predicted synthesizability in the binary Missing composition space, K-means clustering ($k$ = 150) was applied to the t-SNE coordinates of the binary Missing compositions shown in Fig. 6c. Because t-SNE does not preserve global distances, the clustering was used only to delineate visually apparent regions in the projection rather than as a quantitative partition of the latent space. The clusters were ranked by the mean synthesizability score of their constituent compositions, and the chemical family of each of the ten highest-scoring clusters was identified by inspecting the elements and stoichiometries of its compositions (Supplementary Fig. S8). The ten highest-scoring clusters included f-block-containing intermetallic systems, such as f-block elements combined with platinum-group metals or p-block elements, as well as actinide halides, late-transition-metal phosphides, nitrides and polynitrides. Among these clusters, the f-block $AB_2$ intermetallic and f-block + platinum-group metal clusters are discussed in the main text as Regions 1 and 2, respectively (Fig. 6c).

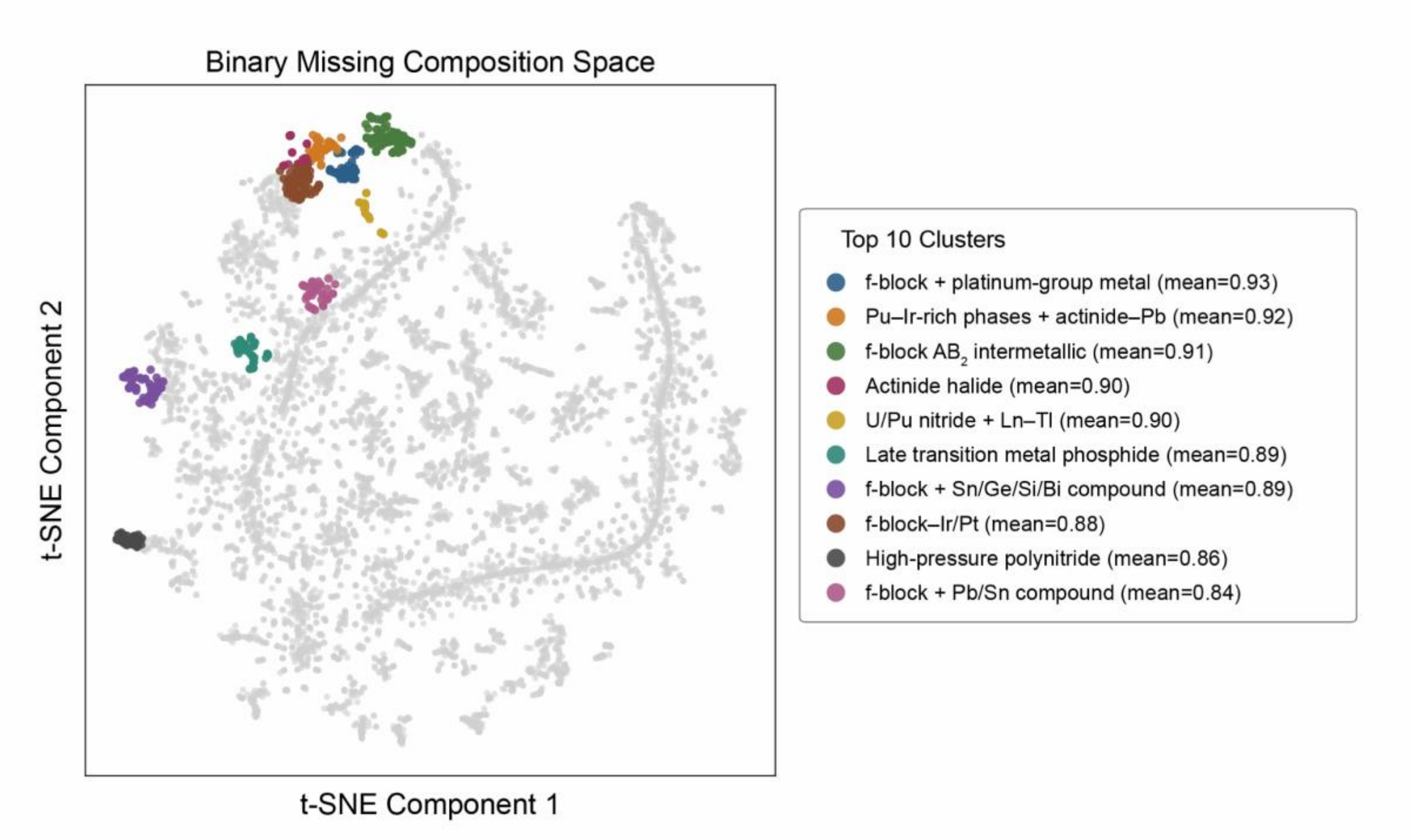

**Supplementary Fig. S8. Top-scoring clusters in the binary Missing composition space.** t-SNE visualization of the binary Missing compositions, with clusters obtained by K-means clustering ($k$ = 150) of the t-SNE coordinates. The ten clusters with the highest mean synthesizability scores are highlighted in color, and all other compositions are shown in gray. The legend lists the characteristic chemical family of each cluster together with its mean synthesizability score. The f-block $AB_2$ intermetallic and f-block + platinum-group metal clusters correspond to Regions 1 and 2 in Fig. 6c, respectively.

## Section S4-2. Chemical Family Analysis of Ternary Missing Compositions

To identify chemically meaningful families among the ternary Missing compositions shown in Fig. 7, we examined which element pairs were statistically enriched among the highest-scoring candidates. We took the 2,000 highest-scoring ternary Missing compositions and compared the occurrence frequency of each element pair against a random sample of 500,000 ternary Missing compositions, using a one-sided Fisher's exact test to rank pairs by enrichment significance (Supplementary Fig. S9a). Each enriched pair was then generalized into a broader chemical family by inspecting the third element(s) most commonly co-occurring with that pair across its highest-scoring members, yielding ten candidate families spanning a range of mean predicted synthesizability scores (Supplementary Fig. S9b).

Among these ten families, Ln–Ln–Tl scored highest overall but was excluded from Fig. 7, because it is closely related to the f-block-containing intermetallic systems already discussed for the binary Missing compositions. Of the remaining families, four (solid electrolyte, polar intermetallics, carbodiimide, and nitridophosphate) were selected for detailed analysis in the main text, corresponding to Regions 1–4 in Fig. 7, respectively.

**a**

Element pair enrichment analysis

| Rank | Element Pair | Count (top 2000) | Count (background) | p-value |
|---|---|---|---|---|
| 1 | (Na, S) | 333 | 437 | $< 10^{-300}$ |
| 2 | (Sb, Zr) | 256 | 510 | $< 10^{-300}$ |
| 3 | (La, Tl) | 124 | 245 | $2\times10^{-199}$ |
| 4 | (Dy, Tl) | 124 | 302 | $3\times10^{-190}$ |
| 5 | (Ce, Tl) | 124 | 308 | $2\times10^{-189}$ |
| 6 | (Bi, Mg) | 102 | 378 | $9\times10^{-141}$ |
| 7 | (Sb, Sc) | 102 | 445 | $2\times10^{-134}$ |
| 8 | (Ca, Pt) | 64 | 602 | $2\times10^{-65}$ |
| 9 | (B, N) | 61 | 943 | $2\times10^{-50}$ |
| 10 | (Mg, Sb) | 38 | 299 | $4\times10^{-42}$ |
| ⋮ | ⋮ | ⋮ | ⋮ | ⋮ |

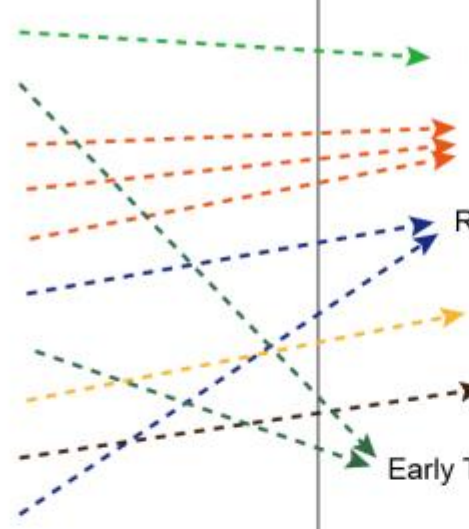

Generalizing into chemical families

| Chemical families |
|---|
| Na + {Si, Ge, P} + S |
| Ln + Ln + Tl |
| RE + Mg + {Sb, Sn, Bi} |
| Ca + Pt + p-block |
| B + Ln + N |
| Early TM + Early TM + Pnictogen |
| ⋮ |

**b**

Top 10 Ternary Missing Chemical Families

| Rank | Family | Definition | Mean Score |
|---|---|---|---|
| 1 | Ln–Ln–Tl | Two distinct lanthanides + Tl | 0.933 |
| 2 | argyrodite | Alkali metal (Cs/Rb/K) + {P, As, Sb, Ge, Si, Sn} + {S, Se} | 0.881 |
| 3 | solid electrolyte | Na + {Si, Ge, P} + S | 0.862 |
| 4 | polar intermetallics | Rare-earth metal + Mg + {Sb, Sn, Bi} | 0.817 |
| 5 | Ca–Pt–p-block | Ca + Pt + {Ge, Ga, Al, In, Sn} | 0.777 |
| 6 | Na–S–X | Na + S + third element other than Si, Ge, P | 0.760 |
| 7 | nitridophosphate | Alkali/alkaline-earth metal + P + N | 0.751 |
| 8 | B–Ln–N | B + lanthanide + N | 0.721 |
| 9 | carbodiimide | Rare-earth metal+ C + N | 0.695 |
| 10 | early TM–pnictide | Two early transition metals (Sc/Ti/Zr/Hf/Ta/V/Nb) + pnictogen | 0.684 |

**Supplementary Fig. S9. Chemical family analysis of ternary Missing compositions. a.** Element pairs enriched among the 2,000 highest-scoring ternary Missing compositions relative to a background set of 500,000 compositions randomly sampled from all ternary Missing compositions. For each element pair, a 2 × 2 contingency table was constructed from the numbers of compositions containing and not containing the pair in the top-scoring and background sets, and *p*-values were computed using a one-sided Fisher's exact test. Enriched element pairs were generalized into broader chemical families by examining their associated third elements, as indicated by the arrows. **b.** The ten highest-scoring chemical families identified through this analysis, together with their compositional definitions and mean synthesizability scores. The solid electrolyte, polar intermetallics, carbodiimide and nitridophosphate families correspond to Regions 1, 2, 3 and 4 in Fig. 7, respectively.

## Section S4-3. Identification of ICSD-Reported Missing Compositions

Although Missing compositions are absent from MP, some of them have experimental reports in ICSD, because MP does not cover all experimentally reported compositions. These compositions, 101 binary and 1,990 ternary, were identified by matching the reduced formulas of Missing compositions against the experimental ICSD entries used in this study, and provide a reference set for evaluating exploration performance that is independent of model predictions. For each composition, we determined whether it was classified as synthesizable by the chronology-free baseline and by ChronoSyn at their calibrated decision thresholds (Supplementary Table S2). ChronoSyn identified a larger fraction of these compositions than the baseline for both binary (37.6% vs 31.7%) and ternary (43.6% vs 39.9%) compositions, with more compositions rescued than lost in both cases. We note that ChronoSyn classified a slightly larger fraction of all ternary Missing compositions as synthesizable than the baseline (22.4% vs 21.1%), which may account for part of the difference observed for ternary compositions, whereas the corresponding fractions for binary compositions were nearly identical (22.6% vs 22.3%). When the decision threshold of ChronoSyn was adjusted to match the fraction of ternary Missing compositions classified as synthesizable by the baseline, ChronoSyn still identified more of the ICSD-reported ternary compositions than the baseline (826 vs 795).

**Supplementary Table S2. Identification of ICSD-reported Missing compositions.** Numbers of Missing compositions with experimental reports in ICSD that were classified as synthesizable by the chronology-free baseline and ChronoSyn trained on the MP-only dataset at their calibrated decision thresholds, with the corresponding fractions in parentheses. Rescued and lost compositions were correctly identified only by ChronoSyn and only by the baseline, respectively.

| Dataset | ICSD-reported | Identified by Baseline | Identified by ChronoSyn | Rescued | Lost |
|---|---|---|---|---|---|
| Binary Missing | 101 | 32 (31.7%) | 38 (37.6%) | 10 | 4 |
| Ternary Missing | 1,990 | 795 (39.9%) | 867 (43.6%) | 190 | 118 |

# Reference

1 Jain, A. *et al.* Commentary: The Materials Project: A materials genome approach to accelerating materials innovation. *APL materials* **1** (2013).

2 Saal, J. E., Kirklin, S., Aykol, M., Meredig, B. & Wolverton, C. Materials design and discovery with high-throughput density functional theory: the open quantum materials database (OQMD). *Jom* **65**, 1501–1509 (2013).

3 Kirklin, S. *et al.* The Open Quantum Materials Database (OQMD): assessing the accuracy of DFT formation energies. *npj Computational Materials* **1**, 15010 (2015).

4 Belsky, A., Hellenbrandt, M., Karen, V. L. & Luksch, P. New developments in the Inorganic Crystal Structure Database (ICSD): accessibility in support of materials research and design. *Structural Science* **58**, 364–369 (2002).

5 Jain, S., White, M. & Radivojac, P. in *Proceedings of the AAAI Conference on Artificial Intelligence.*

6 Jang, J. *et al.* Synthesizability of materials stoichiometry using semi-supervised learning. *Matter* **7**, 2294–2312 (2024).

7 Yoon, Y., Gu, G. H. & Kim, K. Closed-loop workflow of high-entropy materials discovery: efficient and accurate synthesizability prediction via domain-specific local LLMs. *npj Computational Materials* (2026).

8 Pedregosa, F. *et al.* Scikit-learn: Machine learning in Python. *the Journal of machine Learning research* **12**, 2825–2830 (2011).

9 Van der Maaten, L. & Hinton, G. Visualizing data using t-SNE. *Journal of machine learning research* **9** (2008).

10 Poličar, P. G., Stražar, M. & Zupan, B. openTSNE: a modular Python library for t-SNE dimensionality reduction and embedding. *Journal of Statistical Software* **109**, 1–30 (2024).